\documentclass[journal]{IEEEtran}

\ifCLASSINFOpdf
\else
\fi
\usepackage[percent]{overpic} 
\usepackage{float}

\usepackage{cite}
\usepackage{pbox} 
\ifCLASSINFOpdf
\else

\usepackage[dvips]{graphicx}

\DeclareGraphicsExtensions{.eps}
\fi
\usepackage[usenames]{color} 
\usepackage{soul}
\usepackage{times}
\usepackage{amsmath}
\usepackage{epsfig}
\usepackage{colortbl}
\definecolor{light-gray}{gray}{0.95}
\definecolor{gray}{gray}{0.6}
\definecolor{dark-gray}{gray}{0.35}
\usepackage{multirow}
\usepackage[table,dvipsnames]{xcolor}
\usepackage{algorithm,algpseudocode,float}
\usepackage{amssymb}
\usepackage{stfloats}
\usepackage{multirow}
\usepackage[utf8]{inputenc}
\usepackage[english]{babel}

\emergencystretch=\maxdimen
\begin{document}

\title{Optimal Control of SOAs with Artificial Intelligence for Sub-Nanosecond Optical Switching}
%

\author{Christopher~W.~F.~Parsonson*,~\IEEEmembership{Student Member, IEEE,}
        Zacharaya~Shabka*,~\IEEEmembership{Student Member, IEEE,}
        W.~Konrad~Chlupka,~\IEEEmembership{}
        Bawang~Goh,~\IEEEmembership{}
        Georgios~Zervas,~\IEEEmembership{Member, IEEE}
\thanks{This work was supported under the Engineering and Physical Sciences Research Council (EP/R041792/1 and EP/L015455/1), the Industrial Cooperative Awards in Science and Technology (EP/R513143/1), the OptoCloud (EP/T026081/1), and the TRANSNET (EP/R035342/1) grants.}
\thanks{C. W. F. Parsonson, Z. Shabka, W. K. Chlupka, B. Goh, and G. Zervas were all in the Department of Electronic and Electrical Engineering, University College London, London, WC1E 7JE (e-mails: christopher.parsonson.18@ucl.ac.uk, zacharaya.shabka.18@ucl.ac.uk, 
wojciech.chlupka.16@ucl.ac.uk, 
dennis.goh.16@ucl.ac.uk,
g.zervas@ucl.ac.uk).}
\thanks{Manuscript was received on 09/03/2020 and was accepted for publication in the Journal of Lightwave Technology on 21/06/2020.}
\thanks{* Authors made equal contributions.}}

%
%

\markboth{Journal of Lightwave Technology, ~Vol.~X, No.~X, March~2020}%
{Parsonson \MakeLowercase{\textit{et al.}}: iMICIC: Intelligent Multi Impulse Continuous Injection Current as a Generalised Approach to Optimal Sub-Nanosecond Optical Switching}
%



\maketitle

\begin{abstract}
Novel approaches to switching ultra-fast semiconductor optical amplifiers using artificial intelligence algorithms (particle swarm optimisation, ant colony optimisation, and a genetic algorithm) are developed and applied both in simulation and experiment. Effective off-on switching (settling) times of 542 ps are demonstrated with just 4.8\% overshoot, achieving an order of magnitude improvement over previous attempts described in the literature and standard dampening techniques from control theory.  
\end{abstract}


%
\IEEEpeerreviewmaketitle


\section{Introduction} 


%
%
%
%




\IEEEPARstart{B}{y} 2021, annual global data centre network (DCN) traffic will reach $20.6\times 10^{21}$ bytes, 90\% of which will be intra-DCN \cite{Cisco2018}. Additionally, the proportion of requests being serviced by central processing units (CPUs) is expected to decrease from 75\% today to 50\% in 2025 as specialised bandwidth-hungry hardware is installed to enable new machine learning applications \cite{McKinsey2019}. Furthermore, the increasingly common approach of clustering compute resources for large-scale data processing is requiring more network-intensive server-server communication \cite{Andreades2019}. These trends are exerting a growing strain on internal DCNs, in which many of the interconnects are electronic switches. Electronic switches have limited scalability, limited bandwidth, high latency and high power consumption \cite{Zervas2019}, \cite{Wang2018}. As such, switching is presenting a problematic bottleneck for DCN performance, and current network architectures are unfit to meet next-generation DCN requirements.   

Optical switches offer the potential to alleviate many of these network performance issues. With an optical circuit switch (OCS) implementation, there is no packet inspection, buffering, or optical-electrical-optical (OEO) conversion overhead, therefore latency times are significantly lower \cite{Liu2015}. They also have much higher bandwidth, allowing more servers to be connected to the same switch without increasing oversubscription-related buffering, thus improving scalability. Furthermore, the lack of OEO conversion, the transparency to signal modulation format, and the lower heat generation reduces the number of expensive transceiver components needed, the hardware changes required when new transmission protocols are adopted, and the overall network power consumption respectively. The latter is particularly important since networking can account for $>$50\% of the \$20 bn annual DCN power costs, with $CO_2$ emissions equal in volume to the entire aviation industry \cite{Abts2010}. In addition, optical switches have a more compact physical design than their electronic counterparts, allowing for a smaller footprint in DCs.

The difficulty of implementing all-optical DCN switching derives from the bursty nature of most DCN traffic and the lack of an all-optical memory alternative. Since no all-optical memory or processor architectures exist, current DCN packet-switched protocols cannot be implemented with an exclusively optical network architecture based on all-optical switches since header information must be processed and payload information stored on a per-hop basis. An alternative to packet switching is circuit switching, which is possible with an all-optical architecture. However, current state-of-the-art commercial optical switches have slow (100s $\mu$s) switching times. Such long switching times are not compatible with the small data packets that dominate DCN traffic $(90\% < 576 \text{ bytes})$ \cite{Zervas2019} since the switching time would be comparable or greater in size than the forwarding time making for an inefficient network.

For optical circuit switching (OCS) to be compatible with current DCN demands, it must be possible to switch circuits at the packet timescale \cite{Zervas2019}, \cite{Balanici2019}. This requires minimal switching overhead when switching for epochs of the order of 10s-100s of ns.

A promising candidate for realising such a high-speed switch is the semiconductor optical amplifier (SOA). SOAs can be used for either space switching or wavelength switching due to their high and relatively flat optical gain bandwidth. Further benefits of SOAs over other potential optical switching technologies such as MEMS or holograms include fast inherent switching times (theoretically limited only by their $\approx 100$ ps carrier recombination lifetimes \cite{Huang2003}), high extinction/optical contrast ratio, and relatively compact design, making them ideal for low latency-, scalability-, and footprint-constrained DCN applications \cite{Assadihaghi2010}.

The sub-ns off-on time of SOAs allows for an SOA-based optical switch architecture that avoids the issues presented by the lack of all-optical memory/processor alternatives discussed above.  This SOA-based OCS solution is generally more simple and better performing than others suggested by the literature such as optical loop memory \cite{Srivastava2009}, optical burst switching (OBS) \cite{Qiao2004}, \cite{Praveen2005}, \cite{Kiran2007} and hybrid optical packet switching (OPS) \cite{Benjamin2017}, \cite{Wang2018}. However, SOAs have an intrinsic optical overshoot and oscillatory response to electronic drive currents due to exciton density variations and spontaneous emission in the gain region \cite{Paradisi2019}. As demonstrated in this paper, the overshoot and oscillatory optical output result in the key advantage of SOA switching (rapid switching times) being negated, preventing sub-ns switching.

A previous attempt to optimise SOA output applied a `pre-impulse step injection current' (PISIC) driving signal to the SOA \cite{Gallep2002}. This PISIC signal pre-excited carriers in the SOA’s gain region, increasing the charge carrier density and the initial rate of stimulated emission to reduce the 10\% to 90\% rise time from 2 ns to 500 ps. However, this technique only considered rise time when evaluating SOA off-on switching times. A more accurate off-on time is given by the settling time, which is the time taken for the signal to settle within $\pm 5\%$ of the `on' steady state. Before settling, bits experience variable signal to noise ratio, which impacts the bit error rate (BER) and makes the signal unusable until settled, therefore the switch is effectively `off' during this period.

A later paper looked at applying a `multi-impulse step injection current' (MISIC) driving signal to remedy the SOA oscillatory and overshoot behaviour \cite{Figueiredo2015}. As well as a pre-impulse, the MISIC signal included a series of subsequent impulses to balance the oscillations, reducing the rise time to 115 ps and the overshoot by 50\%. However, the method for generating an appropriate pulse format was trial-and-error. Since each SOA has slightly different properties and parasitic elements, the same MISIC format cannot be applied to different SOAs, therefore a different format must be generated through this inefficient manual process for each SOA, of which there will be thousands in a real DC. As such, MISIC is not scalable.  Critically, the MISIC technique did not consider the settling time, therefore the effective off-on switching time was still several ns.

More recent work expands on the driving signal modification shown in \cite{Figueiredo2015}. \cite{reviewer_reference_1} applies the MISIC signal detailed in \cite{Figueiredo2015}, but in addition applies a Wiener filter, where the filter is determined by the steady state value of the SOA response and the mean squared error (MSE) between the output and the filter is minimised by means of finding optimal weight-coefficients of the filter. The work accomplishes a roughly 60\% reduction in guard time, with the goal of reducing guard time as much as possible such that the BER of the output does not exceed a particular level. While the objective of the work (reduce guard time with respect to BER guarantees) is different to that of this work (minimise the settling time of the SOA output), and thus direct comparison is difficult, it is interesting to acknowledge this analogous approach of MSE \& weight optimisation to optimising the output of an SOA.

Similarly, \cite{reviewer_reference_2} explores the optimisation of an SOA by means of both modification of the driving signal and optimisation of the SOA's microwave mounting. A best case of 33\% reduction in guard time is accomplished with an improved microwave mounting architecture and a step driving signal, where various MISIC and PISIC driving signals were also tested. This work demonstrates that significant improvements in guard time can be derived exclusively from improvements being made to the microwave mounting of the SOA - something that is not dealt with in this paper - and that the improvement of the SOA’s output by optimisation of the driving signal does not preclude the simultaneous improvement by optimisation of the microwave mounting. The results therefore are complementary to those presented in this work, which improves the SOA output purely by means of driving signal optimisation. It is speculated here that the optimisation of the SOA's driving signal by the methods presented in this work combined with the optimisation of its microwave mounting could achieve greater improvements in its output than seen in either \cite{reviewer_reference_2} or this work.

The previous solutions discussed so far have had a design flow of first manually coming up with a heuristic for a simplified model of an SOA, followed by meticulous testing and tuning of the heuristic until good real-world performance is achieved. If some aspect of the problem is changed such as the SOA type used or the desired shape of the output signal, this process must be repeated. 

This paper presents a novel and scalable approach to optimising the SOA driving signal in an automated fashion with artificial intelligence (AI) techniques, namely `Particle Swarm Optimisation' (PSO), `Ant Colony Optimisation' (ACO) and `Genetic Algorithms' (GA) \cite{Mata2018}. These algorithms were chosen on the basis that they had previously been applied to proportional-integral-derivative (PID) tuning in control theory \cite{7873803}. Moreover, AI techniques propose the benefit of not requiring prior knowledge of the SOA and therefore provide a means of developing an optimisation method that is generalisable to any SOA-based switch. All algorithms were shown to reduce the settling and rise times to the $O$(100 ps) scale. The algorithms' hyperparameters were tuned in an SOA equivalent circuit (EC) simulation environment and their efficacy was demonstrated in an experimental setup. AI performance was compared to that of step, PISIC and MISIC driving signals as well as the popular raised cosine and PID control approaches to optimising oscillating and overshooting systems, all of which the AI algorithms outperformed. Of the AI algorithms, PSO was found to have both the best performance and generalisability due to the additional hyperparameters and search space restrictions that were required for GA and ACO. All code and plotted data are freely available at \cite{parsonson_shabka_chlupka_2020} and \cite{parsonson_shabka_chlupka_goh_zervas_2020} respectively.

\section{Simulation} 

%
%
%
%





SOAs are typically modelled using simple rate equations. However, as shown in \cite{Ghafouri-Shiraz2004}, the electrical parasitics of an SOA and its surrounding packaging degrade optical signals by broadening the output optical pulse width, reducing the peak optical power (thereby reducing optical contrast), and causing a slight time delay in the emitted optical pulse. Additionally, they alter the relaxation frequency of the SOA output oscillations. As such, modelling the electrical parasitics was crucial to building a simulation environment in which to optimise switching. As described in \cite{Ghafouri-Shiraz2004}, \cite{Figueiredo2011}, and \cite{Tucker1984}, assuming a small circuit model, microwave ECs can be used to more accurately simulate semiconductor diodes by accounting for these electrical parasitics. Therefore, ECs were the chosen approach to SOA modelling for this paper. 

\begin{figure}[!t]
\centering
\includegraphics[scale=0.23]{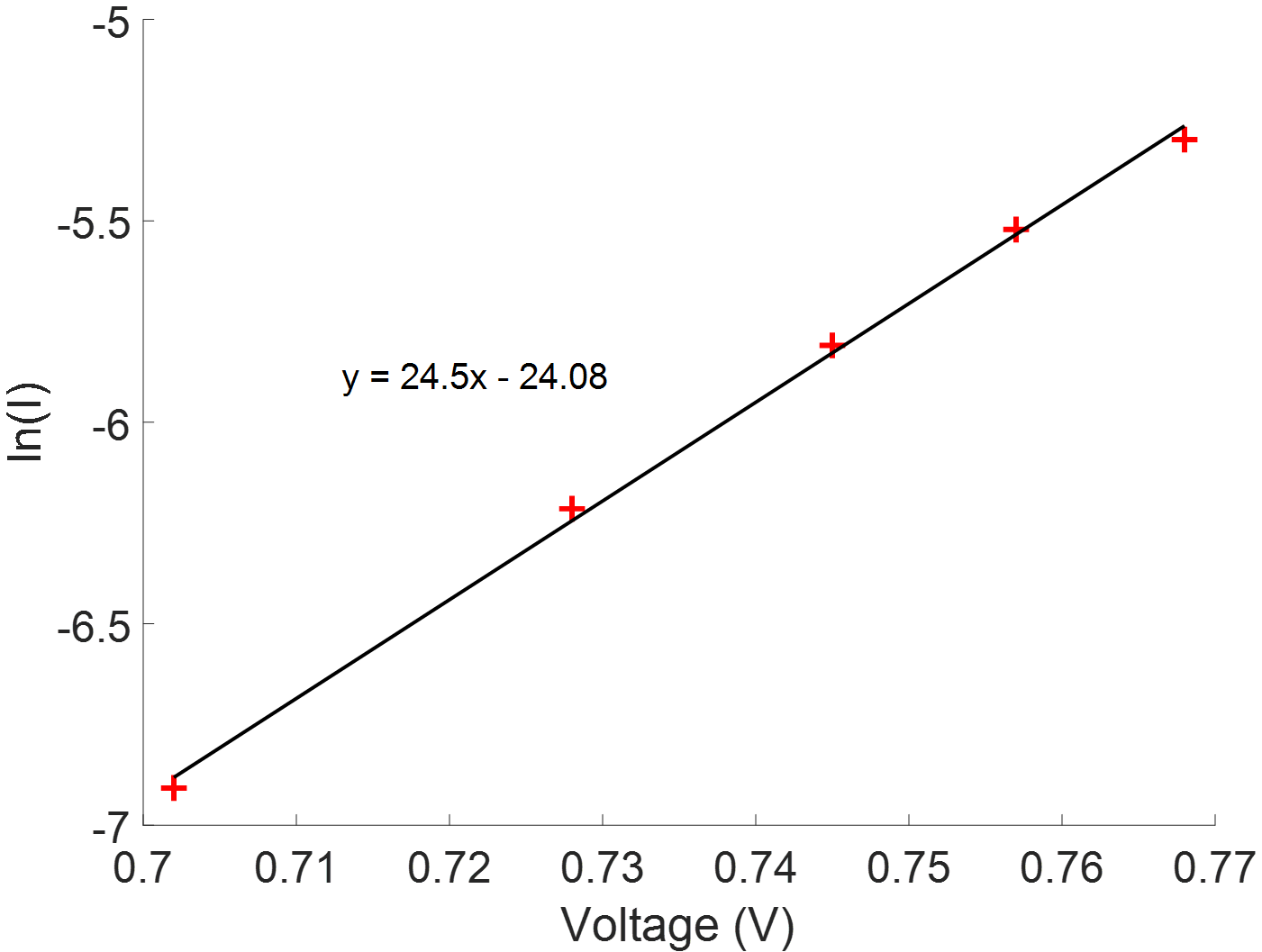}
\caption{Semi-logarithmic I-V plot for the SOA used to calculate $\eta$ and $I_s$.}
\label{fig:soaIVCharacteristics}
\end{figure}

Since at low voltages ($<0.8 V$) the current ($I$) - voltage ($V$) relationship can be described by (\ref{eq:IVRelationship}) ($q$~=~charge, $K_b$~=~Boltzmann constant, $T$~=~temperature), the ideality factor $\eta$ and the saturation current $I_s$ could be calculated as 1.59 and $3.48 \times 10^{-11}$ A respectively using the semi-logarithmic $I$-$V$ curve of the SOA in Fig.~\ref{fig:soaIVCharacteristics}. Using $\eta$, $I_s$ and the internal and external SOA constants taken from the literature for a typical silicon laser diode \cite{Ghafouri-Shiraz2004}, \cite{Figueiredo2011}, the SOA was modelled below the current threshold $I_{TR}$ (2-50 mA) and above $I_{TR}$ (70-110 mA).

\begingroup
\begin{equation} \label{eq:IVRelationship}
ln\big(I\big) = ln\big(I_s\big) + \bigg(\frac{1}{\eta}\bigg) \bigg(\frac{q V}{K_b T}\bigg)
\end{equation}
\endgroup

The SOA in the experimental setup had the optimum trade-off between gain and signal noise at a bias current of 70 mA, therefore the simulated SOA was biased at this current. Using Matlab's Simulink tool, a transfer function (TF) for the SOA EC was obtained and simplified as shown in (\ref{eq:transferFunction}) with the constants defined in Table~\ref{tab:transfer_func_table}. This allowed for custom drive signals to be generated, sent to the biased SOA EC and an optical output measured.

\begin{equation} \label{eq:transferFunction}
\begin{split}
TF                &= \frac{2.01\times 10^{85}} {\sum_{i=0}^9a_is^i }
\end{split}
\end{equation}

\begin{table}[]
    \caption{Constants used in EC transfer function.}
    \label{tab:transfer_func_table}
    \centering
    \tabcolsep 0.1in
    \begin{tabular}{|c|c|c|c|}
             \hline
        $a_9$ & 1.65 & $a_4$ &  $1.37\times 10^{52}$\\
         \hline
        $a_8$ &  $4.56\times 10^{10}$ & $a_3$ & $2.82\times 10^{62}$\\
         \hline
        $a_7$ &  $3.05\times 10^{21}$ & $a_2$ & $9.20\times 10^{71}$\\
         \hline
        $a_6$ & $4.76\times 10^{31}$ &  $a_1$ & $1.69\times 10^{81}$\\
         \hline
        $a_5$ & $1.70\times 10^{42}$ & $a_{0}$ & $2.40\times 10^{90}$ \\
         \hline

    \end{tabular}

\end{table}

In the experimental setup (described later), an arbitrary waveform generator (AWG) with 12 GSPS sampling frequency was used allowing for signal bit windows of 83.3 ps, therefore for 20 ns time periods each signal had 240 points. As such, the optimisation algorithms were searching for a solution in a 240-dimensional search space. Additionally, the oscilloscope had 8-bit resolution, therefore each dimension in the solution could take one of 256 values. The EC simulation environment enabled different driving signals to be rapidly tested.

The constants used for the EC model were taken from the literature (\cite{Ghafouri-Shiraz2004}, \cite{Figueiredo2011}, and \cite{Tucker1984}). The difficulty with SOA modelling, and subsequently also SOA switching, is that there are many variables whose values are difficult to experimentally measure, and which vary significantly even for same-specification SOAs due to parasitics introduced during manufacturing and packaging. Re-measuring these constants for a new SOA would be cumbersome, difficult, and unfruitful since broad assumptions would still need to be made. Furthermore, scaling this bespoke-modelling to 1,000s of SOAs in a single DC would be unrealistic. As such, analytical solutions to SOA switching are not beneficial. Additionally, different driving circuit setups with different amplifiers, bias tees, cabling etc. influence the shape of the driving signal that arrives at the SOA, thereby (if the methods described before this paper are used) requiring more manual tuning every time the equipment surrounding the SOA is changed. This highlights the need for the partially `model-free' AI approaches proposed in this paper, which neither make or require any assumptions about the SOA or the surrounding driving circuit they are optimising, resulting in their optimised driving signals being superior both in terms of performance and scalability relative to traditional analytical and/or manual methods. Here, we borrow the term `model-free' from the field of reinforcement learning, meaning an algorithm that does not initially know anything about the environment in which it must perform its optimisation \cite{10.5555/3312046}.

\begin{figure}[!t]
\centering
\includegraphics[scale=0.4]{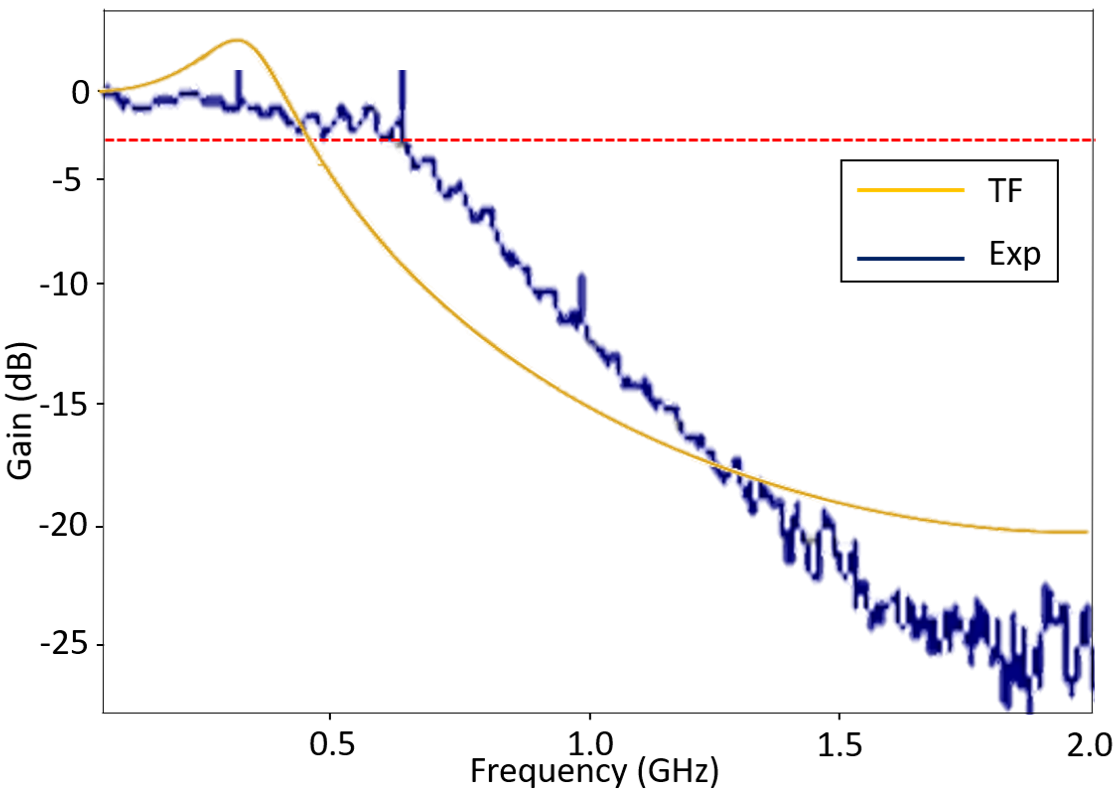}

\caption{Frequency responses of the theoretical transfer function (TF) and the experimental SOA (Exp).}

\label{fig:freq_response}

\end{figure}

Fig.~\ref{fig:freq_response} compares the frequency response of the theoretical TF with the experimental SOA. The TF had a -3dB bandwidth of 0.5 GHz (around 700 ps rise time) compared to the experimental SOA's 0.6 GHz (around 550 ps rise time). These values were similar to one-another and consistent with both the theoretical and experimental optical responses. The differences between the responses were due to the use of EC parameters from the literature which do not exactly match those of our SOA.

\begin{table}[]
    \caption{Factor(s) used on the EC transfer function coefficients to simulate different SOAs (factor = 1 unless stated otherwise).}
    \label{tab:diff_transfer_func_table_of_factors}
    \centering
    \tabcolsep=0.11cm
    \begin{tabular}{|c|c|c|c|c|}
         \hline
        \textbf{TF Component:} & Numerator & $a_0$ & $a_1$ & $a_2$    \\
         \hline
         
        \textbf{Factor(s)}: & 1.0, 1.2, 1.4 & 0.8 & 0.7, 0.8, 1.2 & 1.05, 1.1, 1.2 \\
         \hline
         
    \end{tabular}

\end{table}

\section{Optimisation Algorithms} 

%
%
%
%






All AI algorithms had the goal of minimising the MSE between the actual SOA output and an ideal SOA step output with 0 rise time, settling time and overshoot. The closer the driving signal's corresponding output `process variable' (PV) was to achieving this ideal `set point' (SP), the lower its MSE (defined in (\ref{eq:meanSquaredError})).

\begingroup
\begin{equation} \label{eq:meanSquaredError}
MSE = \frac{1}{m} \sum_{g=0}^{m} \left( PV - SP \right)^2
\end{equation}
\endgroup

\subsection{Particle Swarm Optimisation (PSO)}

\subsubsection{Implementation}

An overview of PSO is given in \cite{Kiranyaz2014}, \cite{Iqbal2015a}. PSO is a population-based AI metaheuristic for optimising continuous nonlinear functions. First proposed in 1995 by \cite{Kennedy1995}, it combines swarm theory by observing natural phenomena such as bird flocks and fish schools with evolutionary programming. In this paper, PSO is adapted to be applicable to SOA drive signal optimisation. 

To apply PSO to SOA optimisation, $n$ particles (driving signals) were initialised at random positions in a hyper-dimensional search space with $m=240$ dimensions (number of points in the signal). Since experimental results showed spurious overshoots after the rising edge and therefore an increase in the settling time, the PSO search space was bounded by a PISIC-shaped `shell' beyond which the particle dimensions could not assume values. An added benefit of the shell was a reduction in the complexity of the problem and therefore also the convergence time. The shell area was a PISIC signal with a leading edge whose width was defined as some fraction of the `on' period of the signal. At each generation, in order to evaluate a given particle position, the MSE in (\ref{eq:meanSquaredError}) was used to calculate the fitness (which was to be minimised). As discussed in \cite{Clerc1999}, the particle inertia weights ($w$) and personal and social cognitive acceleration constants ($c_1$ and $c_2$) can be dynamically adapted as the PSO population evolves. This was done using the update rules in (\ref{eq:updateW}), (\ref{eq:updateM}), and (\ref{eq:updateC}) \cite{Clerc1999} at the start of each generation, where $p_{best_{j}}$ was the historic personal best position of particle $j$, $x_j$ was the position (amplitude taken) of particle $j$, $w(0)$ was the initial inertia weight constant ($0~\leq~w(0)~<~1$), $w(n_t)$ was the final inertia weight constant ($w(0) > w(n_t)$), $m_j(t)$ was the relative fitness improvement of particle $j$ at time $t$, and $c_{max}$ and $c_{min}$ were the maximum and minimum values for the acceleration constants. So long as these values satisfied (\ref{eq:conditionForConvergence}), PSO was guaranteed to converge on some driving signal \cite{VanDenBergh2001}. Using dynamic PSO significantly improved the algorithm's performance (see the `Hyperparameter Tuning' section below).

\begingroup
\begin{equation} \label{eq:updateW}
w_j(t+1) = w(0) + \left[ \left( w(n_t) - w(0) \right) \cdot \left( \frac{e^{m_j(t)} -1} {e^{m_j(t)} + 1} \right) \right]
\end{equation}
\endgroup

\begingroup
\begin{equation} \label{eq:updateM}
m_j(t) = \frac{p_{best_j}(t) - x_j(t)}{p_{best_j}(t) + x_j(t)}
\end{equation}
\endgroup

\begingroup
\begin{equation} \label{eq:updateC}
c_{1,2}(t) = \frac{c_{min} + c_{max}}{2} + \frac{c_{max} - c_{min}}{2} + \frac{e^{-m_j(t)}-1}{e^{-m_j(t)} + 1}
\end{equation}
\endgroup

\begingroup
\begin{equation} \label{eq:conditionForConvergence}
0 \leq \frac{1}{2} \left( c_1 + c_2 \right) - 1 < w < 1
\end{equation}
\endgroup

This PSO process could be repeated until the particles converged on a position with the best fitness (i.e. the optimum SOA driving signal). To help with convergence time and performance, some additional constants were defined:

\begin{itemize}
    \item $iter_{max}$ = Maximum number of iterations that PSO could evolve through before termination. Higher gives more time for convergence but longer total optimisation time.
    
    \item $max\_v\_f$ = Factor controlling the maximum velocity a particle could move with at each iteration. Higher can improve convergence time but, if too high, particles may oscillate around the optimum and never converge.
    
    \item $on\_s\_f$ and $off\_s\_f$ = `On' and `off' suppression factors used to set the minimum and maximum driving signal amplitudes the particle positions could take when the step signal was `on' and `off' respectively. Lower will restrict the particle search space to make the problem tractable for the algorithm, but too low will impact the generalisability of the algorithm to any SOA.
    
    \item $shell\_w\_f$ = Factor by which to multiply the `on' time of the signal to get the width of the leading edge of the PISIC shell. Higher (wider) value will give the algorithm more freedom to rise over a longer period at the leading edge of the signal and improve generalisability, but will also increase the size of the search space and impact convergence.
\end{itemize}


\subsubsection{Hyperparameter Tuning}

The simulation environment enabled the PSO hyperparameters to be rapidly tuned by plotting the PSO learning curve (MSE vs. number of iterations). Since the same PSO algorithm ran multiple times may converge on different minima, each PSO version with its unique hyperparameters was ran 10 times and the 10 corresponding learning curves plotted on the same graph to get a `cost spread' (i.e. how much the converged solution's MSE varied between PSO runs). A lower cost spread gave greater reliability that PSO had converged on the best solution that it could find rather than getting stuck in a local minimum. 

\begin{figure*}[!t]
\centering

    \begin{tabular}{ccc}
    
    \begin{overpic}[width=0.3\textwidth]{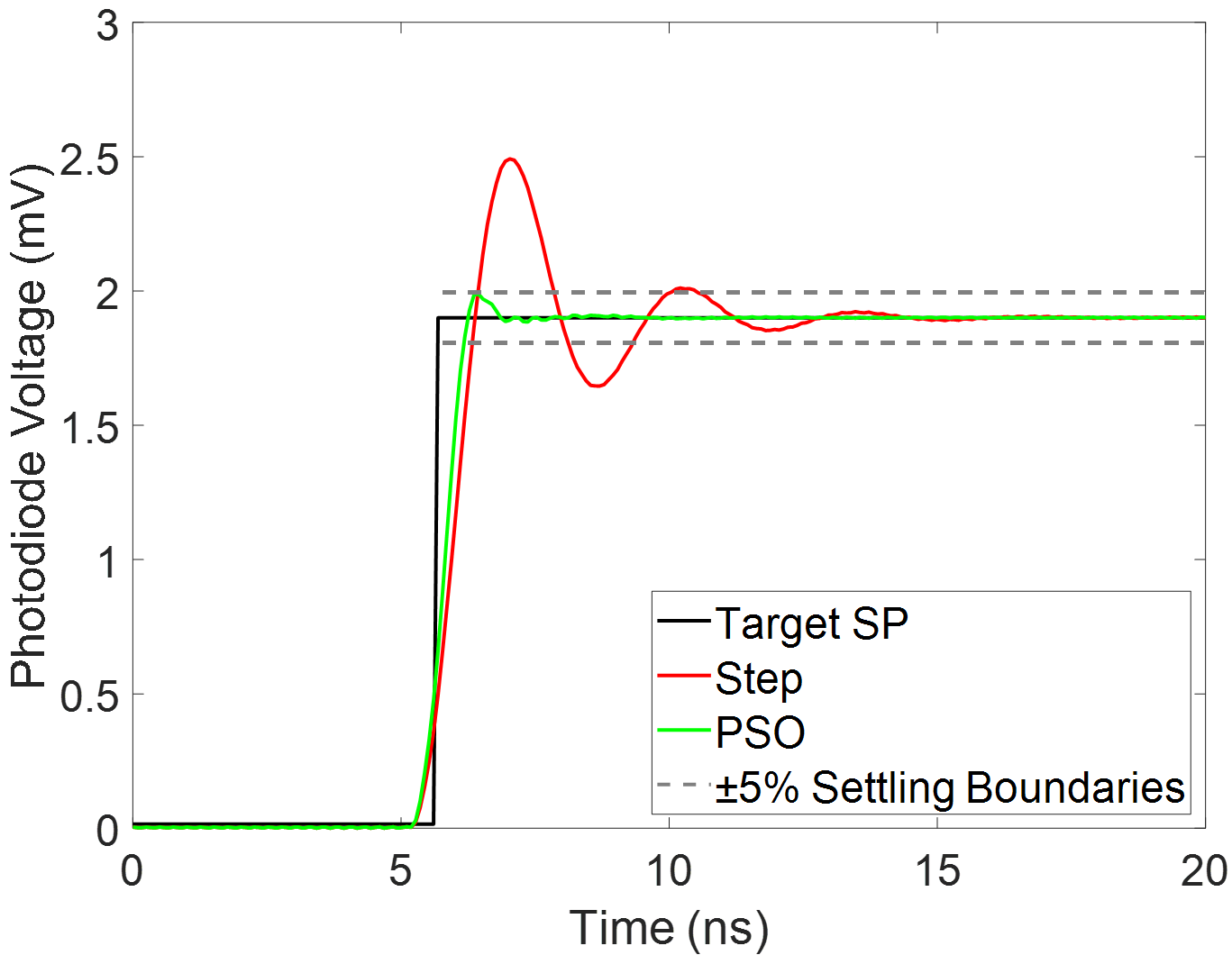}
        \put(15, 68){(a)}
    \end{overpic}
    & 
    \begin{overpic}[width=0.3\textwidth]{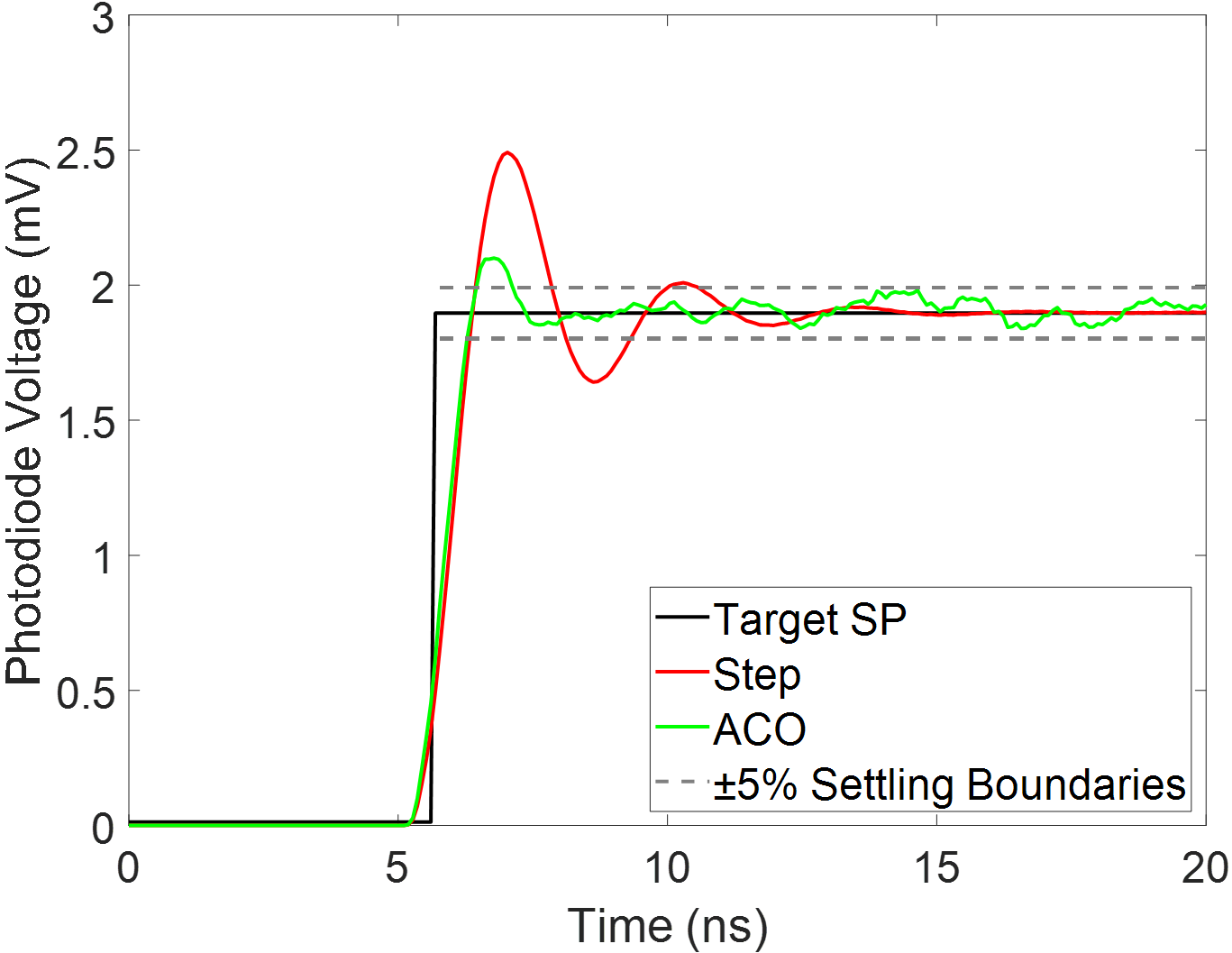}
        \put(15, 68){(b)}
    \end{overpic}
    &
    \begin{overpic}[width=0.3\textwidth]{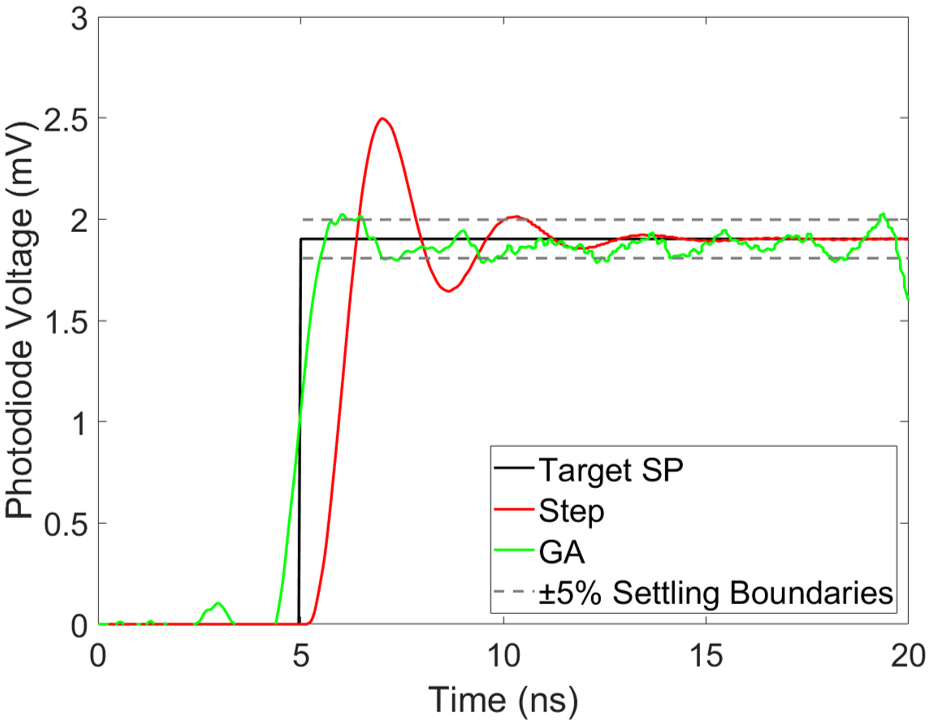}
        \put(15, 68){(c)}
    \end{overpic}
  \\
    \begin{overpic}[width=0.3\textwidth]{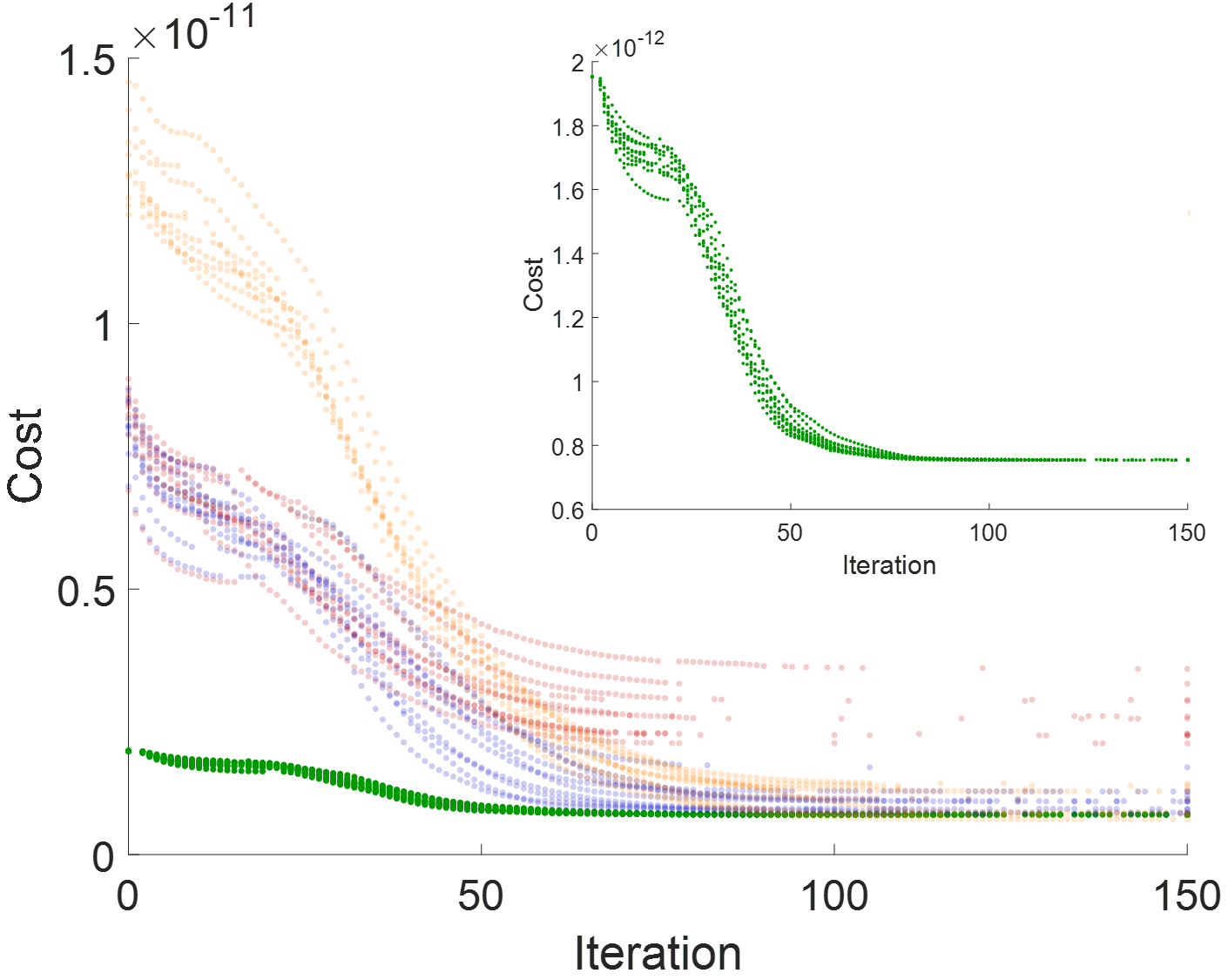} 
        \put(15, 68){(d)}
    \end{overpic}
    &
    \begin{overpic}[width=0.3\textwidth]{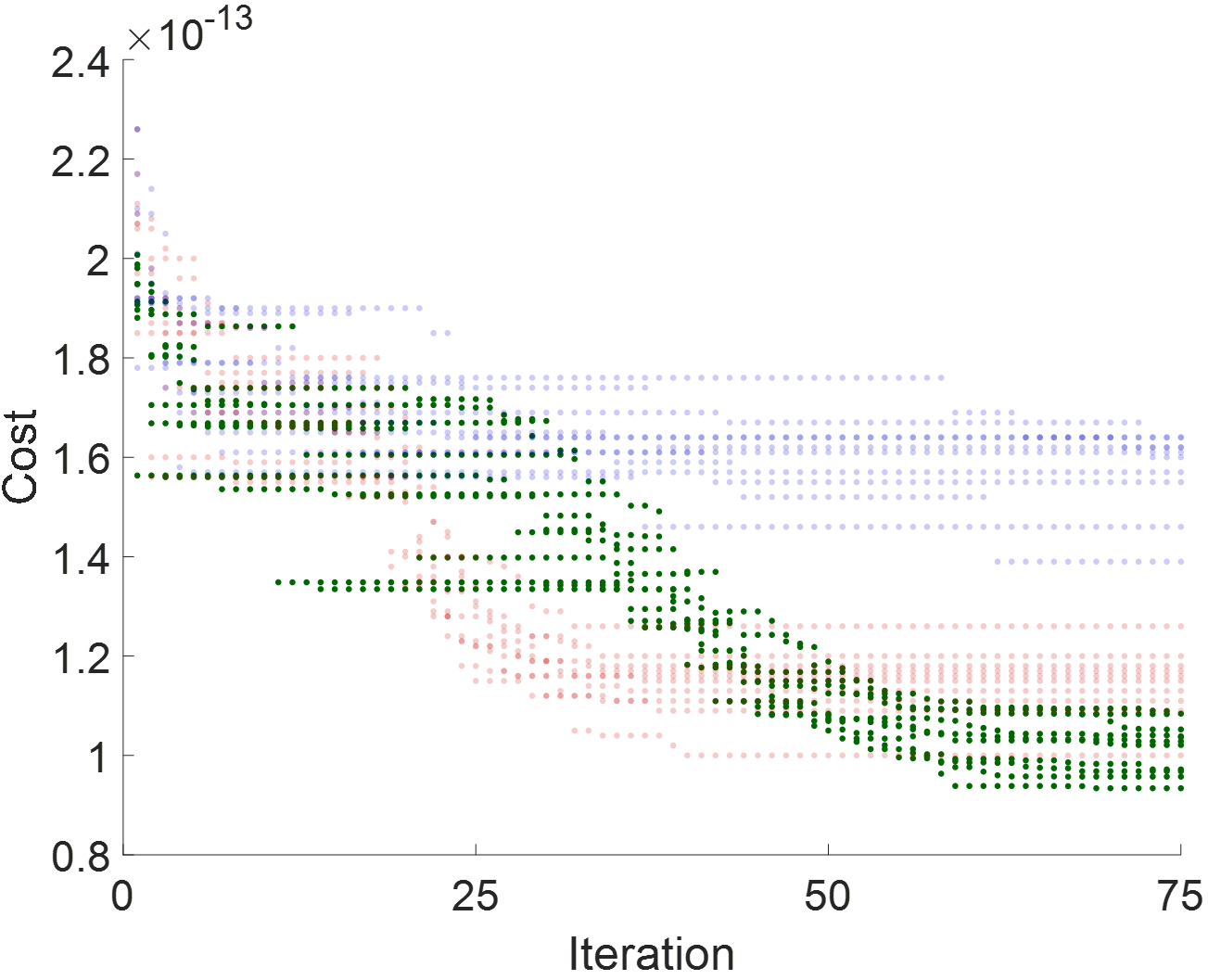}
        \put(15, 68){(e)}
    \end{overpic}
    &
    \begin{overpic}[width=0.3\textwidth]{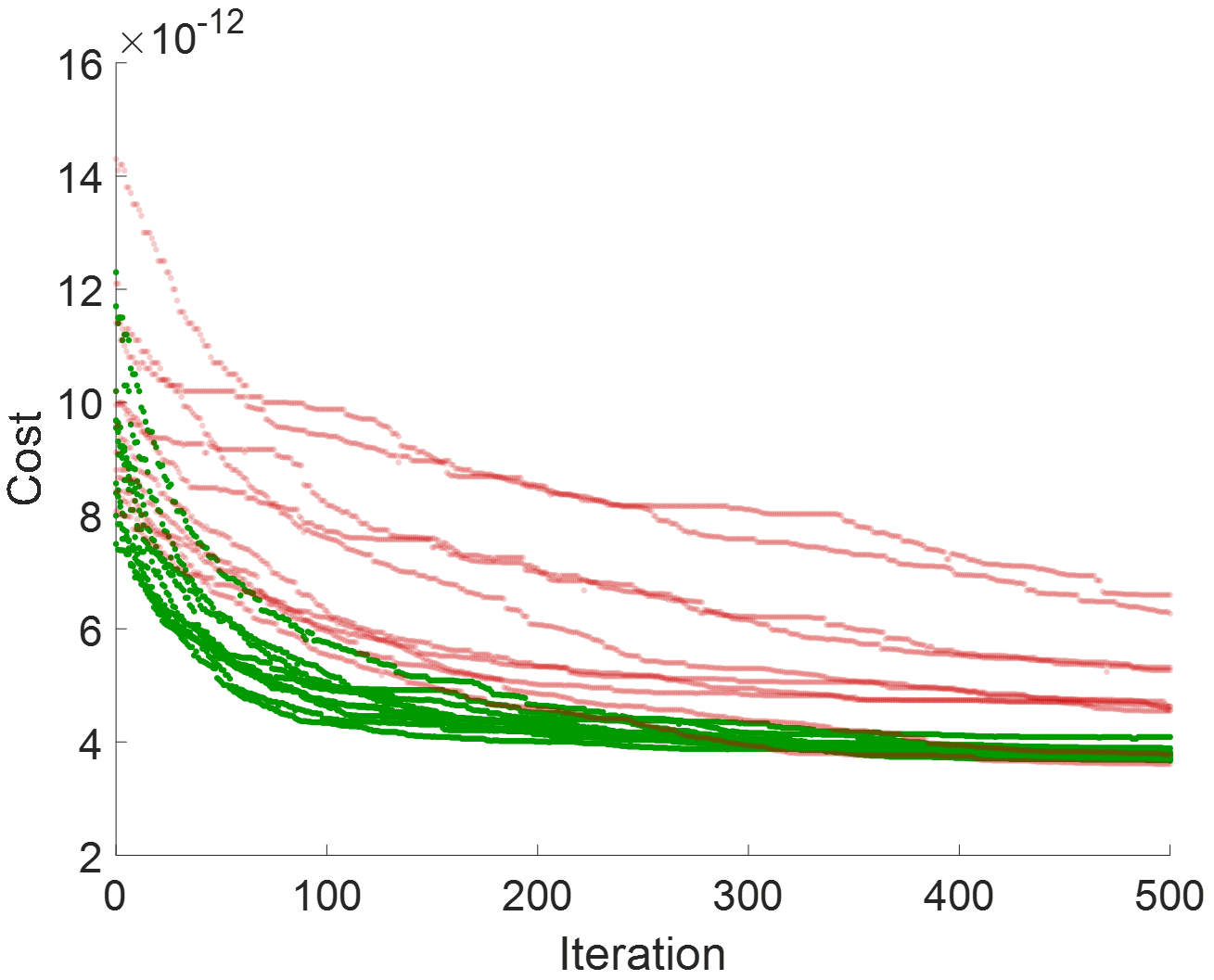}  
        \put(15, 68){(f)}
    \end{overpic}
        
    \end{tabular}

\caption{Simulated SOA optical response to (a) PSO, (b) ACO, and (c) GA driving signals relative to a standard step input. For reference, the target SPs used have also been plotted. Learning curves showing how both the cost spread and the optimum solution improved as the (d) PSO, (e) ACO, and (f) GA algorithms were tuned, showing 10 learning curves for each set of hyperparameters. The curves for the optimum hyperparameters have been plotted in green. For PSO in (d), some additional information has been plotted: i) No dynamic PSO, PISIC shell, or embedded step (red), ii) no PISIC shell or embedded step (blue), iii) no embedded step (orange), and iv) the final PSO algorithm (green, also plotted on separate graph (inserted)). For GA, the i) default DEAP (red) and ii) optimised (green) hyperparameter learning curves have been plotted. For ACO, the blue curve is for a run with a larger pheromone exponent (0.5) value than the optimum, and the red is for a larger dynamic range on the signal search space ($\pm50 \%$).}

\label{fig:simulatedPsoOutputs}

\end{figure*}

To begin with, it was found that using dynamic PSO whereby $w$, $c_1$ and $c_2$ were adapted at the beginning of each generation led to multiple advantages. Firstly, the solution found by 10 dynamic particles had the same MSE as that found by 2,560 static particles, reducing the computation time by a factor of 256. Secondly, the final driving signal found by adaptive PSO was significantly less noisy since it was less prone to local minima. Thirdly, the final MSE found was 63\% lower. Fourthly, although the relative cost spread of dynamic PSO was 72\% compared to 50\% due to the lower MSE, the absolute cost spread was just $8.7\times 10^{-13}$ compared to $140\times 10^{-13}$. Pursuing with dynamic PSO, it was found that placing a `PISIC shell' on the search space (with $shell\_w\_f = 0.1$) beyond which the particles could not travel led to an absolute cost spread of $6.9\times 10^{-13}$ and a further 14\% reduction in the final cost (despite initial costs being higher due to the fact that PISIC signals lead to greater overshoot and subsequently also greater oscillations). Finally, it was also found that initialising one of the $n$ particle positions as a step driving signal improved the convergence time by a factor of two. Using dynamic PSO, a PISIC shell and an embedded step, the following hyperparameter values were found to give the best spread, final cost and convergence time: $iter_{max} = 150$, $n = 160$, $max\_v\_f = 0.05$, $w(0) = 0.9$, $w(n_t) = 0.5$, $c_{min} = 0.1$, $c_{max} = 2.5$, $on\_s\_f = 2.0$, and $off\_s\_f = 0.2$. This final tuning resulted in a cost spread of just 1.8\%. The evolution of this PSO tuning process is summarised in Fig.~\ref{fig:simulatedPsoOutputs}, where the learning curves for the above sets of hyperparameters have been plotted in red, orange, blue and green respectively. The final PSO SOA output, shown in Fig.~\ref{fig:simulatedPsoOutputs}, had a rise time, settling time and overshoot of 669 ps, 669 ps and 3.7\% respectively. Fig.~\ref{fig:simulatedPsoOutputs} also shows the optical response to a step driving signal, showing a rise time, settling time and overshoot of 669 ps, 4.85 ns and 31.1\% respectively. Thus, the simulations indicated that the settling time (and therefore the effective off-on switching time) could be reduced by a factor of 7.2 and the overshoot by a factor of 8.4 compared to a step. Although rise time remained unimproved, the experimental results section shows that, for a real SOA with optical drift, PSO improves all three parameters.

\subsection{Ant Colony Optimisation (ACO)}

\subsubsection{Implementation}

ACO is primarily a path-finding evolutionary algorithm modelled on observations of how ant colonies find food sources in nature. As such, it optimises to find an optimum path along nodes in a graph, $G = \{g_{i}\}$ by means of probabilistic exploration, and colony exploitation across generations of ants. A more comprehensive explanation of ACO can be found in \cite{aco_book}. Of several ACO variants, the `Ant Colony System' algorithm was used in this work.

Since ACO is typically applied to routing problems, considerations must be made as to how to represent parameter selection as such a problem. A system with $N$ parameters each having $M$ possible values can be modelled as a graph with $|N|$ clusters of $|M|$ nodes, where each node maps to a possible value of a particular parameter. A path can then be found that visits one node in each cluster, defining a set of parameter values after each cluster has been visited once and only once.

For example, consider an $N=3$ parameter $(a, b, c)$ system where each parameter can take 1 of $M=2$ possible values which are selected in the order $a\rightarrow b\rightarrow c$. A $(N \times N) \times (M \times M)$ matrix representing the probability of choosing a value for one parameter, given a previous value choice for another, can be written as in (\ref{eq:aco_matrix}) where $\alpha^{xy}_{ij}$ is the probability of choosing value $j$ for parameter $y$, given that value $i$ for parameter $x$ was just chosen. Zeroing the matrix entries appropriately ensures that parameter values are selected in order.

\begin{equation} \label{eq:aco_matrix}
    \begin{pmatrix}
        0 & 0 & \alpha^{ab}_{00} & \alpha^{ab}_{01} & 0 & 0 \\
        0 & 0 & \alpha^{ab}_{10} & \alpha^{ab}_{11} & 0 & 0 \\
        0 & 0 & 0 & 0 & \alpha^{bc}_{00} & \alpha^{bc}_{01} \\
        0 & 0 & 0 & 0 & \alpha^{bc}_{10} & \alpha^{bc}_{11} \\
        0 & 0 & 0 & 0 & 0 & 0 \\
        0 & 0 & 0 & 0 & 0 & 0 \\
    \end{pmatrix}
\end{equation}

\subsubsection{Hyperparameter Tuning}

The important hyperparameters with respect to ACO (specifically the Ant Colony System algorithm used here) are the pheromone exponent (where higher values encourage more exploitation of previously found paths), the evaporation exponent (where higher values discourage exploitation of previously found paths) and the probability of an ant travelling along a randomly selected path. Additionally, the number of ants and generations must be selected.

Parameters were tuned by means of running optimisation routines with one hyperparameter varying across a range of values and the rest kept constant. For each MSE value, the learning curve from 10 different runs were plotted against each other. Just as with PSO, parameter values were selected to prioritise the minimisation of cost spread to ensure that the optimisation technique could give consistent results when used on different occasions. Firstly, it was found that beyond 200 ants, the cost spread did not improve significantly. Similarly, regardless of the spread, the ACO routine was typically converging after between 60 and 75 generations, so a generation cap of 100 was imposed since this was sufficient to guarantee convergence. The values for the other parameters were the pheromone constant $\alpha = 0.25$, the evaporation constant $\rho = 0.5$ and the exploration probability $p = 0.1$. It was also found that minimising the search space by reducing the dynamic range of the signal to $\pm25\%$ centred at $50\%$ of the maximum shortened convergence time without degradation of the final signal, which had the advantage of making matrices memory sizes manageable. No further hyperparameters, such as the PISIC shell applied with the PSO method, were utilised, which is more desirable since fewer hyperparameters simplify the tuning process.

As seen in Fig.~\ref{fig:simulatedPsoOutputs}, the spread of the ACO routine was reduced from 23\% to 14.9\% through tuning, but was still less consistent than the 1.8\% spread of the PSO algorithm. Fig.~\ref{fig:simulatedPsoOutputs} shows the convergence of the Ant Colony System algorithm for various hyperparameter combinations (described in the figure's caption). While the spread in the early iterations of the routine is explained by the embedding of a square signal in the PSO routine described above (since it is very unlikely to randomly initialise a signal better than a square and the ACO does not use any sort of initial signal embedding), the spread in the later stages is thought to be due to some practical limitations of the ACO optimisation method. For $N$ parameters with $M$ values each, the ACO routine requires 2 $(N^2 \times M^2)$ matrices (point-wise multiplied to make a third). A 100 point signal with 100 possible values per point gives a matrix with $100,000,000$ elements. Implemented with the popular NumPy Python library, a minimum of 8 bytes per floating point means such a matrix is on the order of gigabytes. Given the relatively low power PC used in the experiment, restrictions on the state space had to be imposed due to memory limitations. This meant that rather than optimising each point on the signal (240) with the maximum resolution allowed by the AWG (8 bit = 256 points), only 180 points (those in the HIGH state of the initial driving step signal) were optimised with a resolution of 50 points. This means that the state space viewed by the ACO routine was more strongly discretised than that viewed by a method (such as PSO) with lower memory requirements, limiting how optimum the generated signal can be and how well ACO could generalise to other SOAs. Nevertheless, as will be seen, the ACO still produced driving signals that improved upon previous methods. The final ACO tuning output, shown in Fig. \ref{fig:simulatedPsoOutputs}, had a rise time, settling time and overshoot of 753 ps, 1.58 ns and 9.1\% respectively.

\subsection{Genetic Algorithm (GA)}

\subsubsection{Implementation}
GAs are a group of nature-inspired population-based metaheuristics. The term `Genetic Algorithm' relates to the model proposed by John Holland in 1975 \cite{Holland1975}. A detailed explanation of GAs can be found in \cite{Whiteley1994}.

The DEAP Python library \cite{fortin2012deap} was used to implement the canonical GA. Each optimisation started with an initial population of 100 individuals with random positions in order to span as much of the search space as possible. Each individual was represented by an array of 240 points with values within the 7V range, therefore representing a driving signal.

During the evolutionary process, the mutation stage was performed by applying Gaussian noise to some points of each individual. Any individuals with points which went beyond the supported 7V range were discarded.

\subsubsection{Hyperparameter Tuning}

\begin{figure}[!t]
\centering

    \begin{tabular}{cc}
    
    \begin{overpic}[scale=0.17]{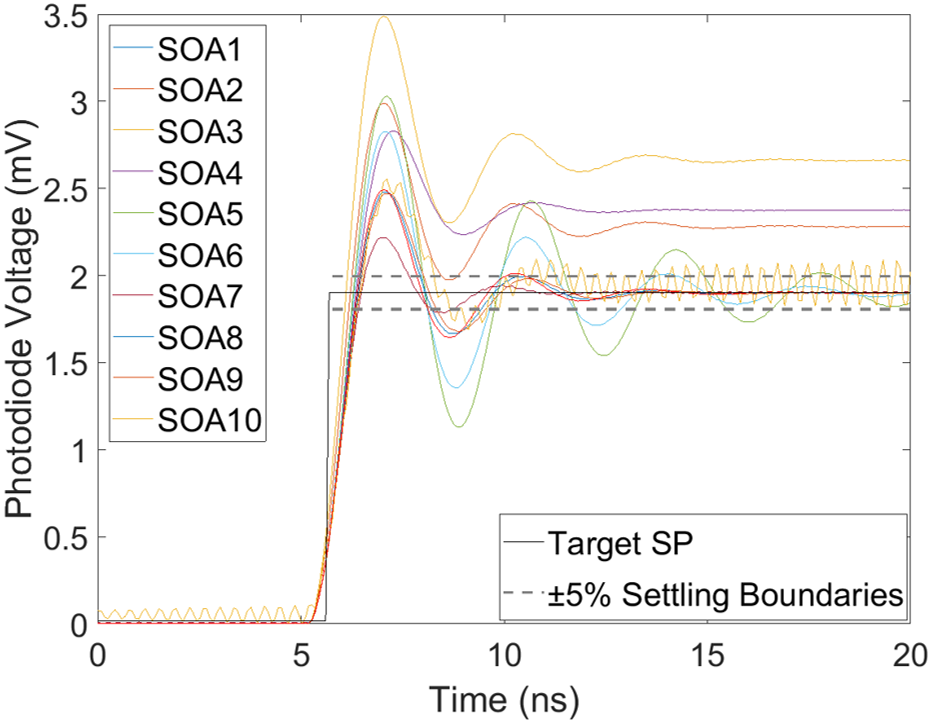}
        \put(87, 69){(a)}
    \end{overpic}
    
    \\
    
    \begin{overpic}[scale=0.17]{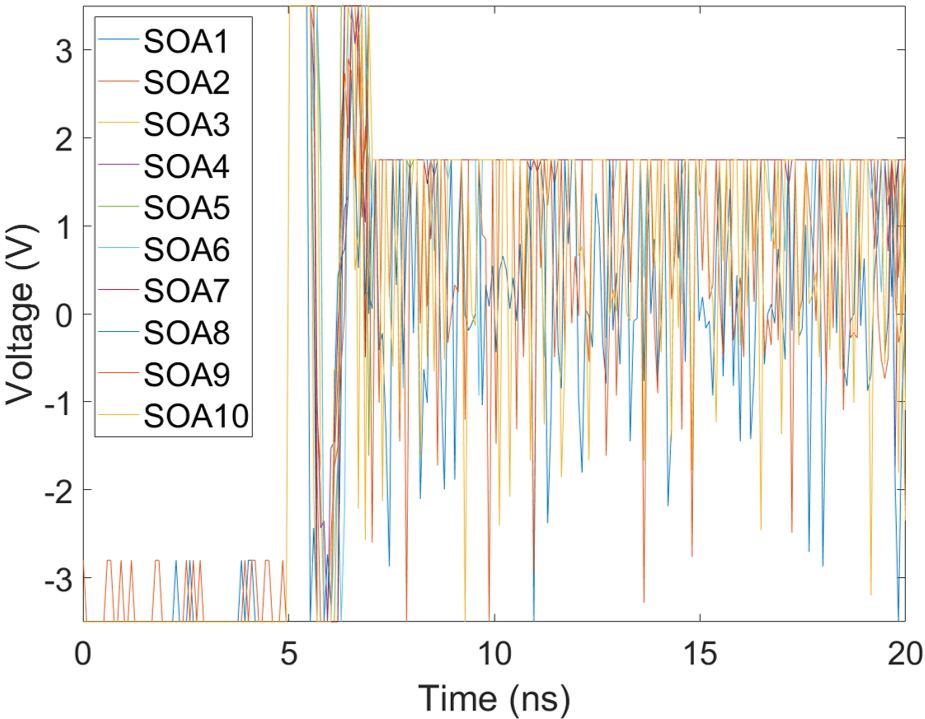}
        \put(87, 71){(b)}
    \end{overpic}
    
    \begin{overpic}[scale=0.17]{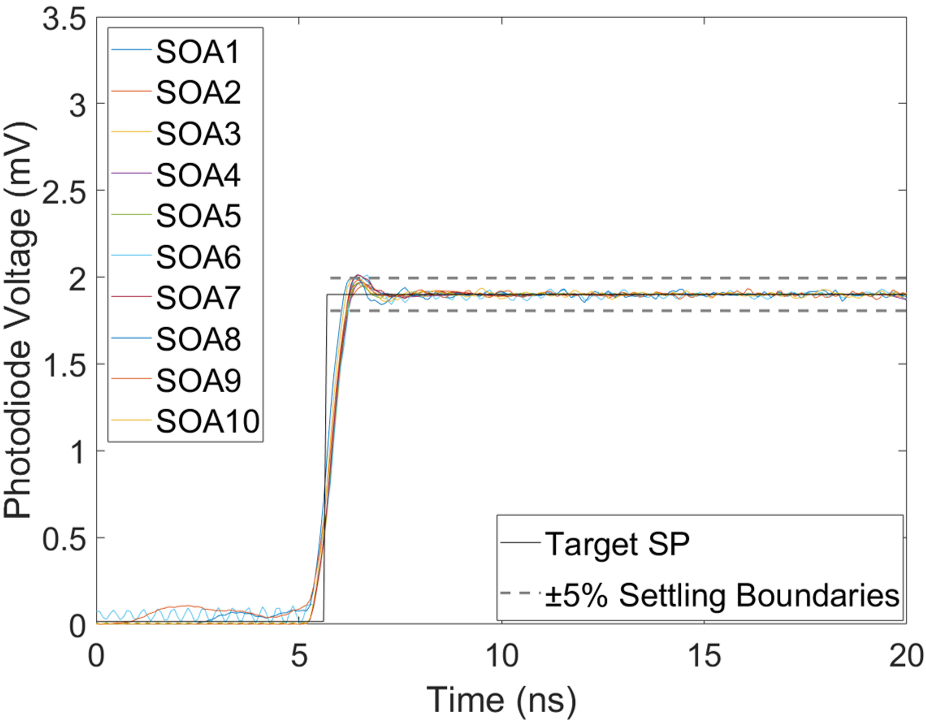}
        \put(87, 69){(c)}
    \end{overpic}
    
    \\
    \begin{overpic}[scale=0.17]{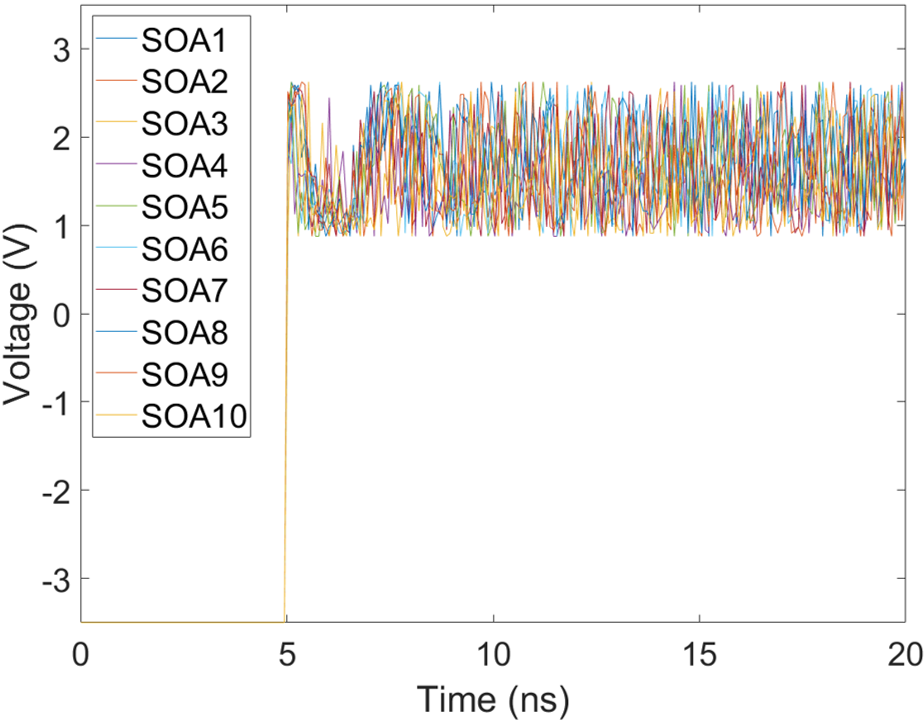}
        \put(87, 71){(d)}
    \end{overpic}
    
    \begin{overpic}[scale=0.17]{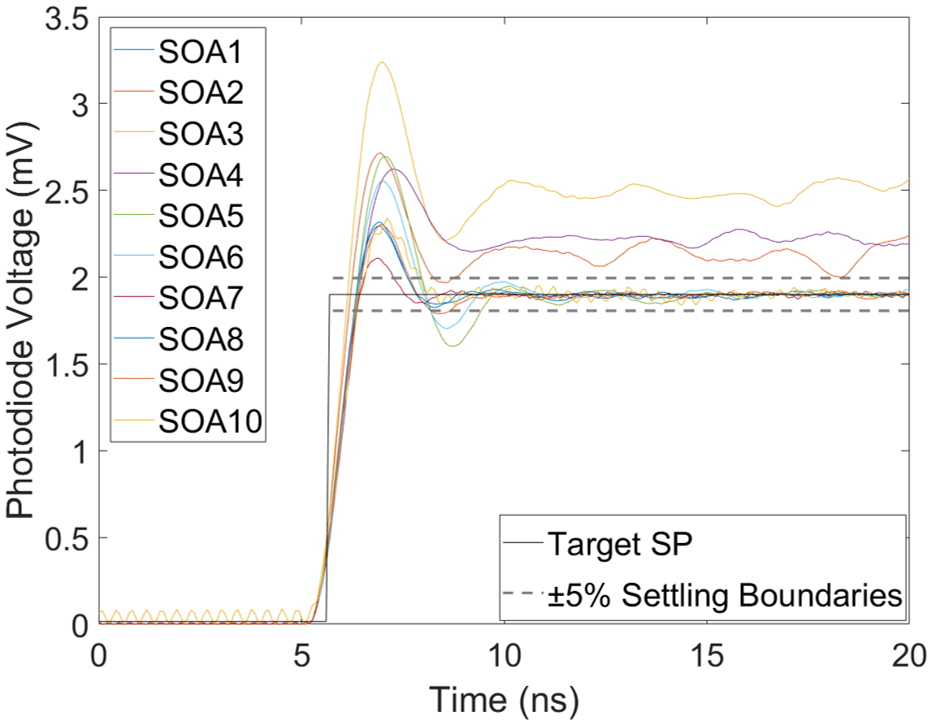}
        \put(87, 69){(e)}
    \end{overpic}
    
    \\
    
    \begin{overpic}[scale=0.17]{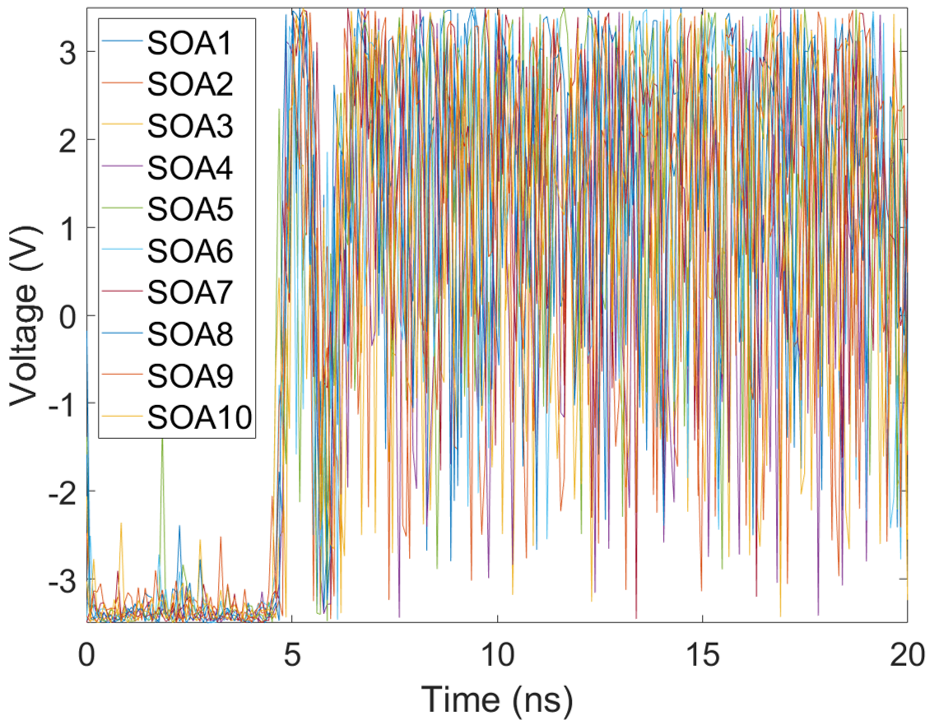}
        \put(87, 69){(f)}
    \end{overpic}
    
    \begin{overpic}[scale=0.17]{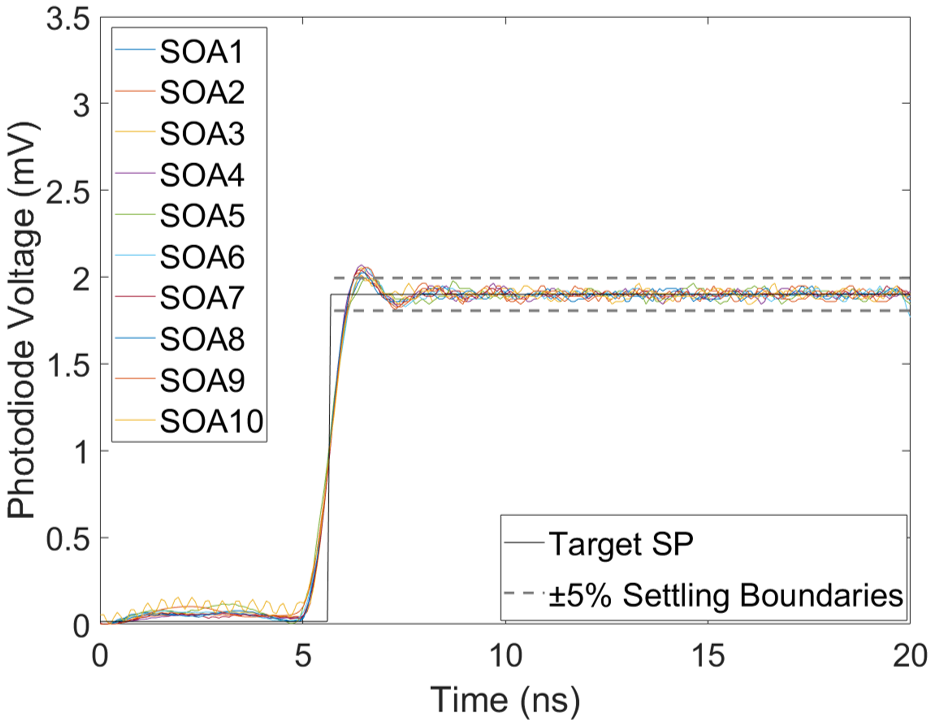}
        \put(87, 69){(g)}
    \end{overpic}

    \end{tabular}

\caption{Simulated SOA optical responses of 10 different SOAs (each with a different transfer function) to (a) step, (c) PSO, (e) ACO, and (g) GA, and the corresponding driving signals for (b) PSO, (d) ACO, and (f) GA. All AI optimisations were done with the same hyperparameters and a common target SP.}

\label{fig:diff_tf_pvs_pso}

\end{figure}

As described in \cite{Whiteley1994}, there are three parts to the evolutionary process: selection, crossover, and mutation. Each of these can be implemented in a few different ways (e.g. Proportionate, Ranking or Tournament for selection \cite{sivaraj2011review}), and each of these implementations use different hyperparameters (e.g. $tournsize$ for Tournament Selection; or $\mu$, $\sigma$, and $indpb$ for Gaussian Mutation). This results in an overall high number of hyperparameters which might significantly impact the probability of the GA getting stuck in a local minimum as well as the speed of convergence. The high number of hyperparameters also meant that there were more values to fine-tune, which made tuning both more complex and time consuming, thereby reducing its generalisability. Since the high number of hyperparameters already impacted generalisability, we refrained from restricting the search space (as done with ACO and with the PSO PISIC shell) to try to still allow for as much generalisability as possible, but this would have the knock-on effect of poorer convergence and a lesser settled signal. However, as demonstrated in Fig. \ref{fig:diff_tf_pvs_pso}, GA was still able to generalise fairly well to 10 different SOAs.

The DEAP library documentation came with a set of suggested default hyperparameter values. These were varied using grid search over 61 optimisations. A limit on the number of generations was set to 500, which was found to be sufficient for convergence.

Mutation was implemented using Gaussian Mutation, which has a probability $indpb$ (mutation rate) of changing each of an individual's points by applying normally distributed noise of mean $\mu$ and standard deviation $\sigma$. Using a negative $\mu$ led to a solution with lower values, while a positive $\mu$ did the opposite - each leading to a lower overall performance, so $\mu$ was set to 0. Decreasing $indpb$ or $\sigma$ slowed down the process as it reduced the overall mutation speed, but increasing either one too much led to the GA getting stuck at local minima. By performing grid search on the hyperparameters, the optimal values were found to be 0.06 and 0.15 respectively. A population size of 60 led to the fastest initial convergence speed (per number of fitness function evaluations), however, the higher number of 100 individuals in a population led to a better overall solution after many generations. Additionally, both $cxpb$ (the probability of mating two individuals), and $mutpb$ (the probability of mutating an individual), were increased significantly from 0.6 to 0.9 and from 0.05 to 0.3 respectively. Increasing $tournsize$ (which controls the number of randomly selected individuals from which to choose the best one for the next generation \cite{miller1995genetic}) above 4 did not have an impact on the convergence, whereas using the values of 2 and 3 significantly slowed down the process. Most hyperparameters did not change by much from the DEAP library's default values since the initial values were almost optimal and changing them led to a slower convergence.

Fig.~\ref{fig:simulatedPsoOutputs} shows the 10 learning curves for the default hyper parameters (red) and the optimised parameters (green), where the cost spread was reduced from 58.6\% to 10.8\%. Fig.~\ref{fig:simulatedPsoOutputs} also shows the simulated SOA output of the tuned GA algorithm with a rise time, settling time and overshoot of 799 ps, 2.55 ns, and 9.0\% respectively.

The hyperparameters of the AI algorithms can be used to address the general problem of `SOA optimisation'. This is because the hyperparameters are only for restricting the search space to reduce the size of the problem, and restricting how much the algorithm can change its solution between iterations; they are specific to the general SOA optimisation problem, but \textit{not} to a specific SOA. The EC simulation environment provided a useful test bed in which to tune the algorithm hyperparameters and allow optimisation of any SOA (even though drive signal solutions derived from simulations are not directly transferable to experiment).

To test the above claim that these algorithms can in theory be generalised to any SOA, we generate 10 different transfer functions each modelling a different SOA. These were generated by multiplying the coefficients in Table~\ref{tab:transfer_func_table} by various factors (summarised in Table~\ref{tab:diff_transfer_func_table_of_factors} so as to be reproducible), thereby simulating SOAs with different characteristics. The optical outputs of these different SOAs in response to the same step driving signal are shown in Fig.~\ref{fig:diff_tf_pvs_pso}. Using the PSO and GA algorithms with the \textit{same hyperparameters}, all 10 of these SOAs are able to be optimised with no changes to the algorithms, as shown in Fig.~\ref{fig:diff_tf_pvs_pso} (where the AI electrical drive signals have been included for reference). Due to search space restrictions, ACO could not generalise. For all 10 SOAs, a common target set point was chosen. The set point was defined as a perfect 0 overshoot, rise time and settling time step response based on the steady states of the initial step response of one of the simulated SOA's. However, the target can be arbitrarily defined by the user if a different optical response is required, demonstrating the flexibility of the AI algorithms to optimise optical outputs with respect to specific problem requirements. Relative to this target set point, the performances are summarised in Table~\ref{tab:diff_transfer_func_comparison_table}. Signals that did not settle have been marked as `-' and excluded from performance summary metrics. PSO had the greatest generalisability to optimising the settling times of different SOAs. Researchers in our field should therefore be able to black box our PSO AI approach and optimise their SOAs even though they will have different equivalent circuit components from the specific device(s) optimised in this paper.

\begin{table}[]
\footnotesize
    \caption{Performance summary for the techniques applied to the 10 different simulated SOAs, given in the format min | max | mean | standard deviation (best in bold).}
    \begin{itemize}
        \item \scriptsize{Signals marked `-' never settled.}
    \end{itemize}
    \label{tab:diff_transfer_func_comparison_table}
    \centering
    \tabcolsep=0.11cm
    \begin{tabular}{|c|c|c|c|}
         \hline
        {\textbf{Technique}} & \pbox{20cm}{\textbf{Rise Time (ps)}} & {\textbf{Settling Time (ns)}} & {\textbf{Overshoot (\%)}} \\
         \hline
         
         Step & \scriptsize{\textbf{502}, \textbf{753}, 653, 86.4} & \scriptsize{3.1, -, 5.8, 3.0} & \scriptsize{16.5, 70.4, 39.2, 14.1} \\
         \hline
         
         PSO & \scriptsize{669, 837, 703, \textbf{58.5}} & \scriptsize{\textbf{0.67}, \textbf{1.3}, \textbf{0.87}, \textbf{0.20}} & \scriptsize{\textbf{2.51}, \textbf{6.01}, \textbf{4.46}, \textbf{1.22}} \\
         \hline
         
         ACO & \scriptsize{\textbf{502}, \textbf{753}, \textbf{644}, 79.4} & \scriptsize{1.6, -, 2.6, 0.82} & \scriptsize{11.1, 70.4, 32.6, 17.0} \\
         \hline

         GA & \scriptsize{760, 930, 793, \textbf{58.5}} & \scriptsize{1.0, 1.5, 1.3, 1.5} & \scriptsize{4.31, 9.36, 7.04, 1.54} \\
         \hline

    \end{tabular}

\end{table}

\section{Experimental Setup} 

%
%
%
%





\begin{figure}[!t]
\centering
\includegraphics[scale=0.35]{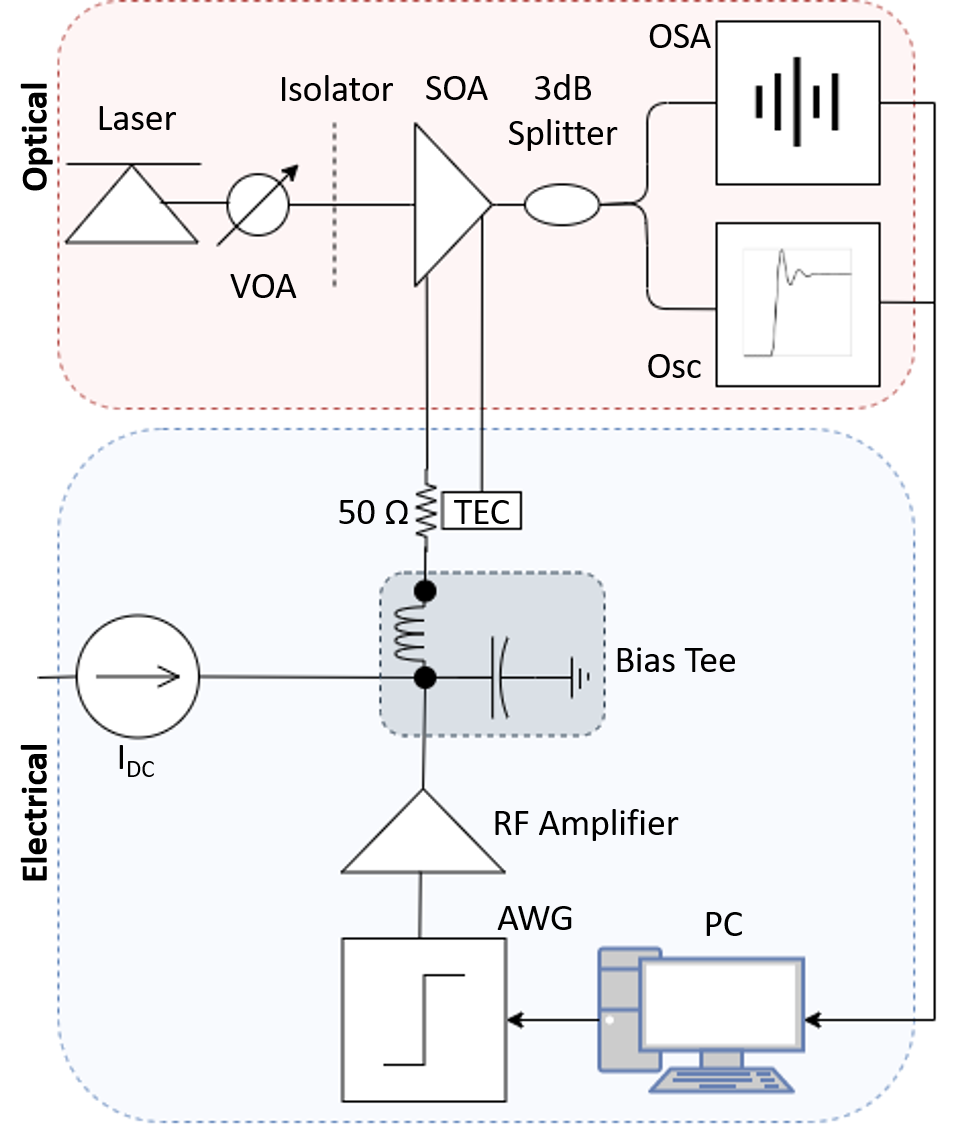}
\caption{Diagram of the SOA experimental setup used.}
\label{fig:experimentalSetup}
\end{figure}

The experimental setup is shown in Fig.~\ref{fig:experimentalSetup}. An INPhenix-IPSAD1513C-5113 SOA with a 3dB bandwidth of 69 nm, a small signal gain of 20.8 dB, a 0-140 mA bias current range, a saturation output power of 10 dBm, a response frequency of 0.6 GHz, and a noise figure of 7.0 dB was used. An SHF 100 BP RF amplifier was selected by calculating the amplified MSE relative to the direct signal for different amplifiers, enabling a full dynamic range peak-to-peak voltage of 7V. A $50\Omega$ resistor was placed before the SOA, allowing for the maximum allowed dynamic current range of 140 mA to be applied across the SOA.

The 70 mA optimum SOA bias current was found by measuring how MSE, optical signal-to-noise ratio (OSNR), rise time, overshoot, and optical gain varied with current. A 70 mA bias using a -2.5 dBm SOA input laser power produced the lowest rise time and MSE. The SOA was therefore driven between 0 and 140 mA centred at 70 mA. The other equipment used included a Lightwave 7900b lasing system, an Agilent 8156A optical attenuator, an LDX-3200 Series bias current source, a Tektronix 7122B AWG with 12 GSPS sampling frequency, an Anritsu M59740A optical spectrum analyser (OSA), and an Agilent 86100C oscilloscope (OSC) with an embedded photodiode. The RF signal going into the SOA had a rise time of 180 ps, therefore this was the best possible rise time (and settling time) that the SOA could have achieved. Throughout the experiments, a wavelength of 1,545 nm was used.

\section{Experimental Results} 

\begin{table*}[h]
\begin{center}
    \tabcolsep 0.1in
    \caption{
        Comparison of SOA Optimisation Techniques. (Best in bold).}
        \begin{itemize}
            \item \scriptsize $^{a}$ Though exact value not reported in \cite{reviewer_reference_2}, it is referred to as being `below 500 ps'.
            \item \scriptsize $^{b}$ Comparison of the ASM mounting against the commercial STF mounting.
            \item \scriptsize $^{c}$ Exact value not reported in \cite{reviewer_reference_2} so percentage improvement is (approximately) inferred from a graph presented in \cite{reviewer_reference_2}. Comparison made at bias current value corresponding to the best case performance of the best performing ASM mount + drive combination and is compared against the STF mount + drive at the same bias and for the same drive (step was best performing in the reported metrics).
            \item \scriptsize $^{d}$ Comparison is made between the best and worst cases presented in \cite{reviewer_reference_1}.
            \item \scriptsize $^{e}$ Several variants of the `MISIC' format were tested in \cite{Figueiredo2015} and the best is used here for comparison.
            \item \scriptsize $^{f}$ Comparison made with respect to the performance of the STEP driving signal presented in \cite{Figueiredo2015}.
        \end{itemize}
    
    \begin{tabular}{|c|c|c|c|c|c|}
        \hline
        \multicolumn{1}{|p{3cm}|}{\centering \textbf{Method \\ (Technique)}}
        & \textbf{Reference}
        & \multicolumn{1}{|p{2.2cm}|}{\centering \textbf{Rise Time, ps \\ (Reduction, \%)}}
        & \multicolumn{1}{|p{2.2cm}|}{\centering \textbf{Settling Time, ps \\ (Reduction, \%)}}
        & \multicolumn{1}{|p{2.2cm}|}{\centering \textbf{Overshoot, \% \\ (Reduction, \%)}}
        & \multicolumn{1}{|p{2.2cm}|}{\centering \textbf{Guard Time, ps \\ (Reduction, \%)}}
        \\\hline
        \multicolumn{1}{|p{3cm}|}{\centering \textbf{PSO \\ (Signal Optimisation)}}
        & This work 
        & \multicolumn{1}{|p{2.2cm}|}{\centering 454 ps \\ (35\%)} 
        & \multicolumn{1}{|p{2.2cm}|}{\centering \textbf{547 ps \\ (85\%)}} 
        & \multicolumn{1}{|p{2.2cm}|}{\centering 5\% \\ -} 
        & \multicolumn{1}{|p{2.2cm}|}{\centering - \\ -} 
        \\\hline
        \multicolumn{1}{|p{3cm}|}{\centering \textbf{ACO \\ (Signal Optimisation)}}
        & This work 
        & \multicolumn{1}{|p{2.2cm}|}{\centering 413 ps \\ (41\%)} 
        & \multicolumn{1}{|p{2.2cm}|}{\centering 560 ps \\ (85\%)} 
        & \multicolumn{1}{|p{2.2cm}|}{\centering 4.8\% \\ -}
        & \multicolumn{1}{|p{2.2cm}|}{\centering - \\ -} 
        \\\hline
        \multicolumn{1}{|p{3cm}|}{\centering \textbf{GA \\ (Signal Optimisation)}}
        & This work 
        & \multicolumn{1}{|p{2.2cm}|}{\centering \textbf{340 ps \\ (51\%)}} 
        & \multicolumn{1}{|p{2.2cm}|}{\centering 825 ps \\ (78\%)} 
        & \multicolumn{1}{|p{2.2cm}|}{\centering 10.3\% \\ -} 
        & \multicolumn{1}{|p{2.2cm}|}{\centering - \\ -} 
        \\\hline
        \multicolumn{1}{|p{3cm}|}{\centering \textbf{PISIC \\ (Signal Optimisation)}}
        & This work 
        & \multicolumn{1}{|p{2.2cm}|}{\centering 502 ps \\ (28\%)} 
        & \multicolumn{1}{|p{2.2cm}|}{\centering 4350 ps \\ (-17\%)} 
        & \multicolumn{1}{|p{2.2cm}|}{\centering 40.5\% \\ -} 
        & \multicolumn{1}{|p{2.2cm}|}{\centering - \\ -} 
        \\\hline
        \multicolumn{1}{|p{3cm}|}{\centering \textbf{MISIC1 \\ (Signal Optimisation)}}
        & This work 
        & \multicolumn{1}{|p{2.2cm}|}{\centering 502 ps \\ (28\%)} 
        & \multicolumn{1}{|p{2.2cm}|}{\centering 4020 ps \\ (-8\%)} 
        & \multicolumn{1}{|p{2.2cm}|}{\centering undershot \\ -} 
        & \multicolumn{1}{|p{2.2cm}|}{\centering - \\ -} 
        \\\hline
        \multicolumn{1}{|p{3cm}|}{\centering \textbf{Raised Cosine \\ (Signal Optimisation)}}
        & This work 
        & \multicolumn{1}{|p{2.2cm}|}{\centering 921 ps \\ (-32\%)} 
        & \multicolumn{1}{|p{2.2cm}|}{\centering 4690 ps \\ (-26\%)} 
        & \multicolumn{1}{|p{2.2cm}|}{\centering undershot \\ -} 
        & \multicolumn{1}{|p{2.2cm}|}{\centering - \\ -} 
        \\\hline
        \multicolumn{1}{|p{3cm}|}{\centering \textbf{PID Control \\ (Signal Optimisation)}}
        & This work 
        & \multicolumn{1}{|p{2.2cm}|}{\centering 501 ps \\ (28\%)} 
        & \multicolumn{1}{|p{2.2cm}|}{\centering 4020 ps \\ (-8\%)} 
        & \multicolumn{1}{|p{2.2cm}|}{\centering 2.3\% \\ -} 
        & \multicolumn{1}{|p{2.2cm}|}{\centering - \\ -} 
        \\\hline
        \multicolumn{1}{|p{3cm}|}{\centering \textbf{ASM Mounting + STEP Drive \\ (Microwave Mounting Optimisation)}} 
        & \cite{reviewer_reference_2} 
        & \multicolumn{1}{|p{2.2cm}|}{\centering - \\ -} 
        & \multicolumn{1}{|p{2.2cm}|}{\centering - \\ -} 
        & \multicolumn{1}{|p{2.2cm}|}{\centering \textbf{$\approx$ 5\% $^{[c]}$ \\ ($\approx$ 75\% $^{[b,c]}$)}}
        & \multicolumn{1}{|p{2.2cm}|}{\centering $\approx$ 500 ps $^{[a]}$ \\ ($\approx$ 33\% $^{[b,c]}$)} 
        \\\hline
        
        \multicolumn{1}{|p{3cm}|}{\centering \textbf{STEP Drive + Wiener Filtering \\ (Signal Optimisation + Filtering)}} 
        & \cite{reviewer_reference_1} 
        & \multicolumn{1}{|p{2.2cm}|}{\centering - \\ -} 
        & \multicolumn{1}{|p{2.2cm}|}{\centering - \\ -} 
        & \multicolumn{1}{|p{2.2cm}|}{\centering - \\ -}
        & \multicolumn{1}{|p{2.2cm}|}{\centering \textbf{286 ps} \\ \textbf{(60\% $^{[d]}$)}} 
        \\\hline
        
        \multicolumn{1}{|p{3cm}|}{\centering \textbf{PISIC Drive \\ (Signal Optimisation)}} 
        & \cite{Figueiredo2015} 
        & \multicolumn{1}{|p{2.2cm}|}{\centering 115 ps \\ (34\% $^{[f]}$)} 
        & \multicolumn{1}{|p{2.2cm}|}{\centering - \\ -} 
        & \multicolumn{1}{|p{2.2cm}|}{\centering 25\% \\ (-56\% $^{[f]}$)}
        & \multicolumn{1}{|p{2.2cm}|}{\centering - \\ -} 
        \\\hline
        \multicolumn{1}{|p{3cm}|}{\centering \textbf{MISIC-6 Drive $^{[e]}$ \\ (Signal Optimisation)}} 
        & \cite{Figueiredo2015} 
        & \multicolumn{1}{|p{2.2cm}|}{\centering 115 ps \\ (34\% $^{[f]}$)} 
        & \multicolumn{1}{|p{2.2cm}|}{\centering - \\ -} 
        & \multicolumn{1}{|p{2.2cm}|}{\centering 12.5\% \\ (22\% $^{[f]}$)}
        & \multicolumn{1}{|p{2.2cm}|}{\centering - \\ -} 
        \\\hline
    \end{tabular}
    \label{tab:review_comparison_table}
    \end{center}
\end{table*}
%
%
%
%




\begin{figure}[!t]
\centering
\includegraphics[scale=0.26]{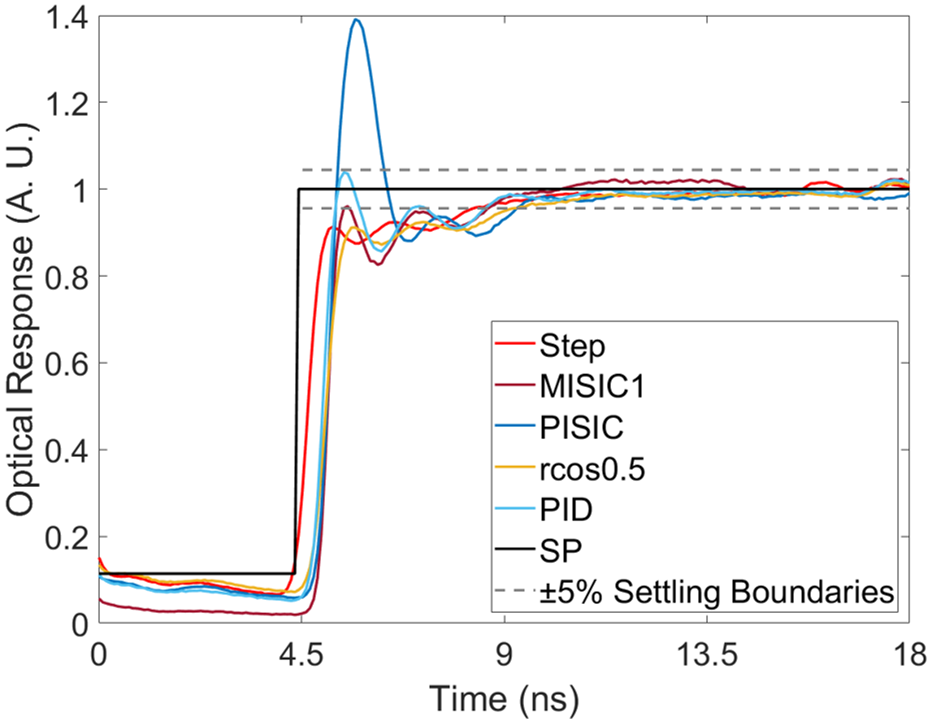}
\caption{Experimental SOA responses to the step, PISIC, MISIC1, raised cosine and PID driving signals.}
\label{fig:expSOAResponses}
\end{figure}

In this section the experimental results for the SOA responses to step, PISIC, MISIC, raised cosine, PID and AI driving signals have been compared. The objective was to reduce the off-on switching time and power oscillations (measured by the settling time and overshoot metrics).

A step driving signal was the simplest format used to drive the SOA. Fig.~\ref{fig:expSOAResponses} (which  has  been  normalised  with  respect  to  the steady  state  value  as done in \cite{Figueiredo2015} for easy comparison) shows the SOA optical response to a step driving signal, resulting in a rise time, settling time and overshoot of 697 ps, 3.72 ns and 0.0\% (since it undershot the steady state) respectively. 

The PISIC format proposed by \cite{Gallep2002} was applied to the SOA with 2.95V step + 4.05V impulse, and the response is shown in Fig.~\ref{fig:expSOAResponses} with a rise time, settling time and overshoot of 502 ps, 4.35 ns and 40.5\% respectively. The form of the PISIC pulse used was optimised for the SOA in use, where different step-impulse voltage combinations (as done in [18]) were tested, as well as varying widths of the pre-impulse section of the PISIC signal as a percentage of the total signal length centered at the percentage used in [18]. It was found that a 500ps pulse width gave the best results.

The MISIC 1-6 bit-sequences proposed by \cite{Figueiredo2015} were applied with 2.95V step + 4.05V impulse, where the same step-impulse voltage combinations were tested as for PISIC. The format with the best performance was MISIC1, whose response is shown in Fig.~\ref{fig:expSOAResponses} with a rise time, settling time and overshoot of 502 ps, 4.02 ns and 0.0\% (undershot) respectively. 

A popular approach to optimising oscillating systems in control theory is the raised cosine approach, whereby the rising step for a signal of period $T$ is adapted to a rising cosine defined by the frequency-domain piecewise function in (\ref{eq:raisedCosine}). As $\beta$ increases ($0 \leq \beta \leq 1$), the rate of signal rise decreases. The best performing raised cosine was $\beta = 0.5$, whose response is shown in Fig.~\ref{fig:expSOAResponses} and whose rise time, settling time and overshoot were 921 ps, 4.69 ns and 0.0\% (undershot) respectively. 

\begingroup
\small
\begin{equation} \label{eq:raisedCosine}
H(f) = \begin{cases}
    1, & \text{if $f \leq \frac{1-\beta}{2T}$} \\
    \frac{1}{2} \left[ 1 + cos \left( \frac{\pi T}{\beta} \left[f - \frac{1-\beta}{2T} \right] \right) \right], & \text{if $ \frac{1-\beta}{2T} < f \leq \frac{1+\beta}{2T} $} \\
    0, & \text{otherwise}
    \end{cases}
\end{equation}
\endgroup

Another popular approach in control theory is the PID controller. The optical response of the PID control signal is shown in Fig.~\ref{fig:expSOAResponses}, with a rise time, settling time and overshoot of 501 ps, 4.02 ns and 2.3\% respectively. In order to quickly obtain values for the 3 PID parameters, $K_c, K_i \text{ and } K_d$, a First Order Plus Dead Time (FOPDT) model was applied to the SOA, where the key parameters for this model ($K_p, \tau_p \text{ and } \theta_p$) can be measured directly from the step response of the device. The PID tuning parameter, $\tau_c$, which is inversely proportional to the magnitude of the response to offset, was tested with values between that of an `aggressive' tuning regime $(\tau_c \approx 0.1)$ and a `conservative' one $(\tau_c \approx 10.0)$. The results shown in Fig.~\ref{fig:expSOAResponses} are with $\tau_c = 5.0$ which was found to be the best performing value.

\begin{figure*}[!t]
\centering

    \begin{tabular}{ccc}
    \begin{overpic}[width=0.3\textwidth]{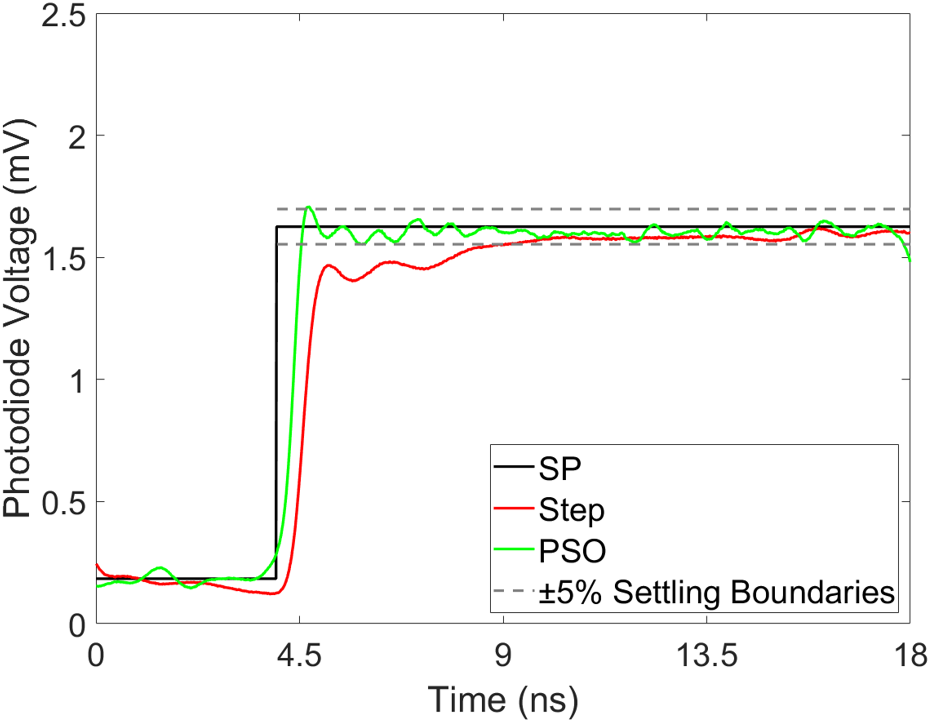} 
        \put(15, 68){(a)}
    \end{overpic}
    &
    \begin{overpic}[width=0.3\textwidth]{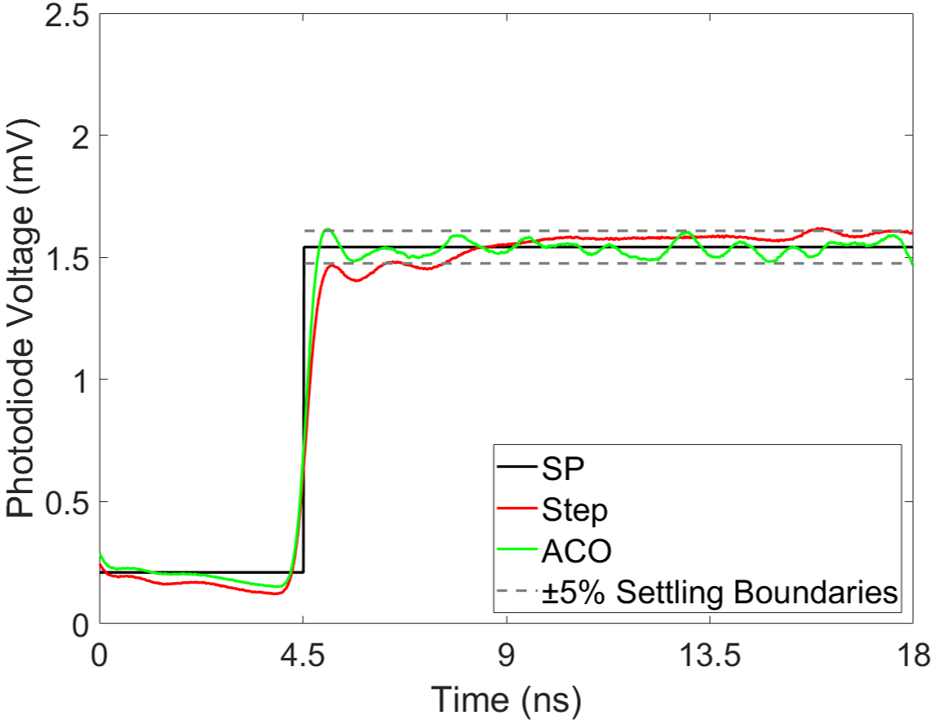}
        \put(15, 68){(b)}
    \end{overpic}
    &
    \begin{overpic}[width=0.3\textwidth]{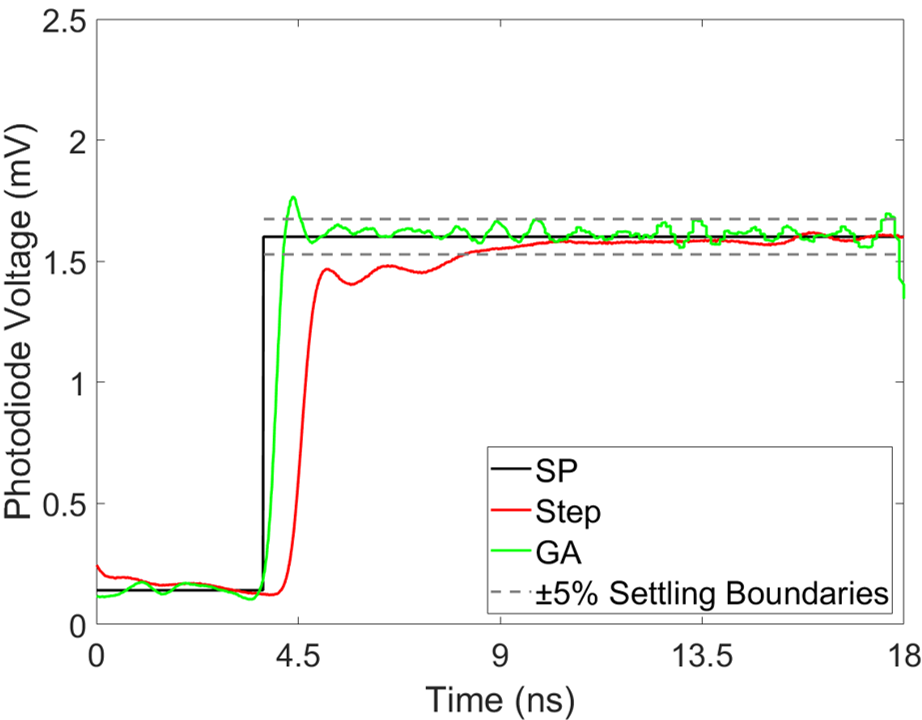}  
        \put(15, 68){(c)}
    \end{overpic}

    \end{tabular}

\caption{Experimental results showing the optimised SOA optical outputs for (a) PSO, (b) ACO, and (c) GA.}
\label{fig:experimentalResults}
\end{figure*}

The PSO algorithm used in the simulation environment was applied to the real SOA. The SP and the PSO response are shown in Fig.~\ref{fig:experimentalResults}, with a rise time, settling time and overshoot of 454 ps, 547 ps and 5.0\% respectively.

\begin{figure*}[!t]
\centering

    \begin{tabular}{ccc}
    \begin{overpic}[width=0.3\textwidth]{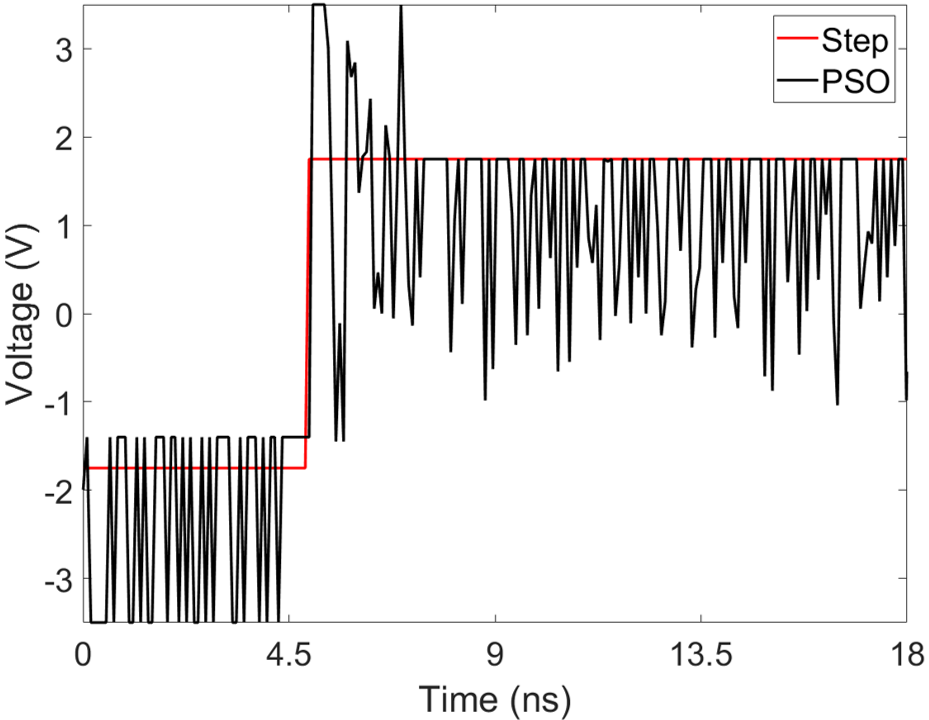} 
        \put(15, 68){(a)}
    \end{overpic}
    &
    \begin{overpic}[width=0.3\textwidth]{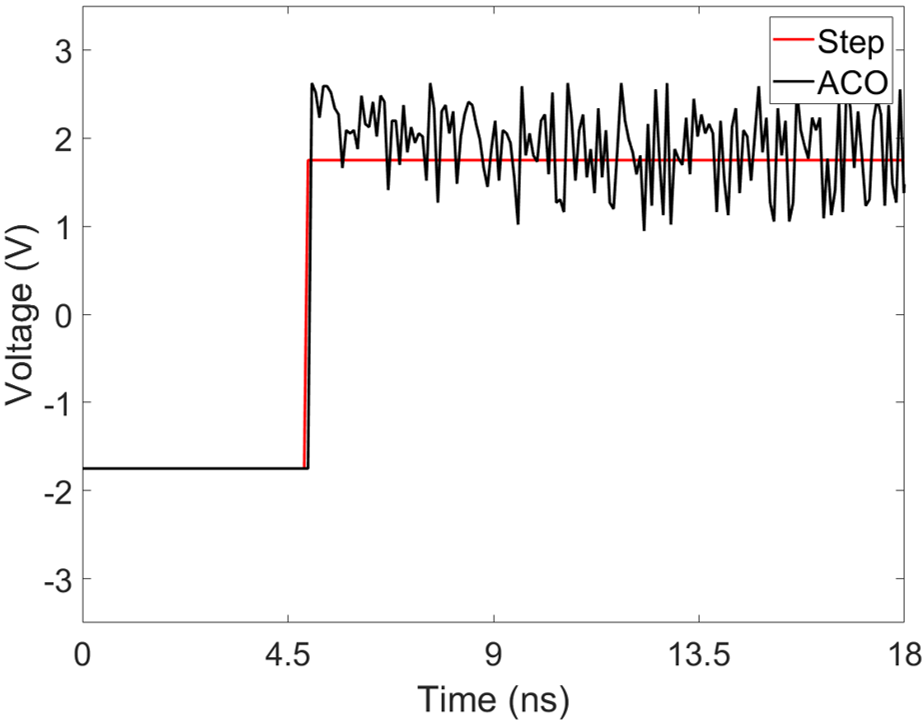}
        \put(15, 68){(b)}
    \end{overpic}
    &
    \begin{overpic}[width=0.3\textwidth]{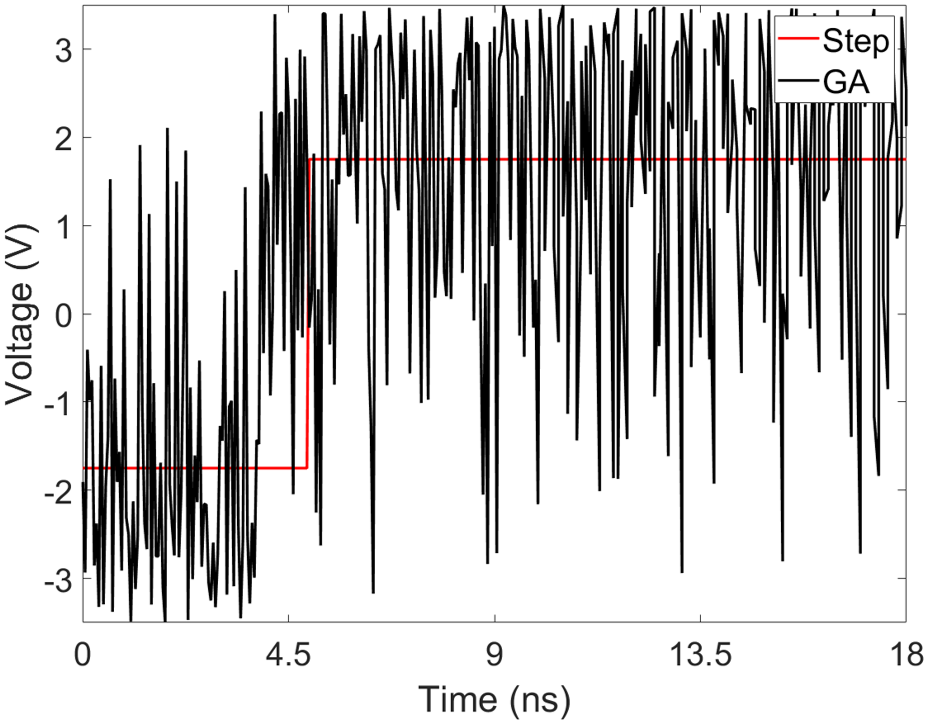}  
        \put(15, 68){(c)}
    \end{overpic}

    \end{tabular}

\caption{Experimental results showing the optimised SOA electrical driving signal inputs for (a) PSO, (b) ACO, and (c) GA.}
\label{fig:ai_optimised_ops}
\end{figure*}








An ACO run with 200 ants accomplished a rise time, settling time and overshoot of 413 ps, 560 ps and  4.8\% respectively, performing similarly well to the PSO algorithm. The ACO result is shown in Fig.~\ref{fig:experimentalResults}

\begin{figure}[!t]
\centering
\includegraphics[scale=0.26]{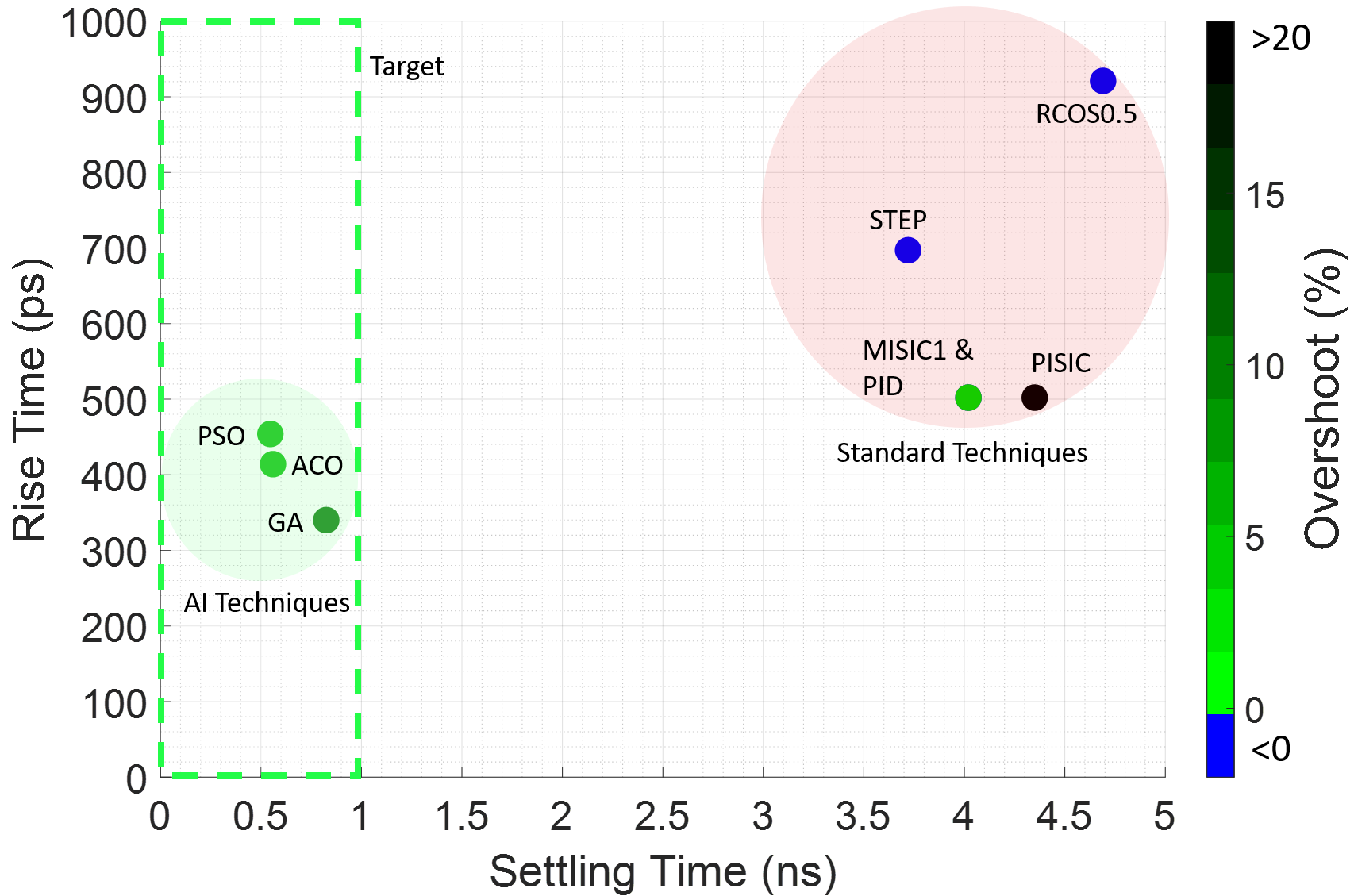}
\caption{Scatter plot comparing the experimental rise times, settling times and overshoots of all the driving signals tested. The outlined target region highlights the performance required for truly sub-nanosecond optical switching.}
\label{fig:Standard_3D_Scatter}
\end{figure}

Similarly, the GA result shown in Fig.~\ref{fig:experimentalResults} had a rise time, settling time, and overshoot of 340 ps, 825 ps, and 10.3\% respectively. The rise times of the AI algorithms were an order of magnitude improvement on the step's, and the settling times (and therefore the effective off-on switching time) were several factors faster than the previous MISIC1 optimum from the literature, bringing SOA switching times truly down to the hundred ps scale. A scatter plot comparing these data is shown in Fig.~\ref{fig:Standard_3D_Scatter}. 

By comparison, PSO had the lowest settling time and therefore the lowest overall switch time. We hypothesise that this was due to the fact that PSO, being less memory-hungry than ACO and having superior convergence properties compared to GA as a result of having fewer hyperparameters to fine-tune and a smaller search space with the PISIC shell, was able to be given a better search space-hyperparameter tuning trade-off, and therefore was able to find a more optimum driving signal. This larger search space also enabled PSO to explore a wider variety of drive signal solutions without needing a large number of hyperparameters tuned (which adds complexity), allowing PSO to generalise to a more diverse set of SOAs than either ACO or GA were able to. Therefore, although in theory all AI algorithms used were powerful and generalisable, due to the number of hyperparameters and search space restrictions that were required in practice, PSO had both the best performance and generalisability, although GA came close to matching PSO.

Table \ref{tab:review_comparison_table} shows results (both absolute, and relative improvement for cross-comparison) of the rise time, settling time, overshoot and guard time for all methods implemented in this work, as well as a variety from the literature. The rows associated with \cite{Figueiredo2015} are the results for the optimised PISIC and MISIC-6 signals defined and implemented in this work. This is distinguished from the other two columns with `PISIC' and `MISIC' methods referred to as coming from this work, which are a re-implementation of the methods described in \cite{Figueiredo2015} but applied to and optimised for a different experimental setup.

Finally, Fig.~\ref{fig:ai_optimised_ops} shows the electrical drive signals found by each algorithm. Whilst we stress that the main focus of this paper is the \textit{method} rather than the \textit{specific drive signal}, the drive signal is important for real-World implementation and general understanding of the search space restrictions used. As Fig.~\ref{fig:ai_optimised_ops} shows, the derived driving signals are noisy despite a smooth resultant optical output. This is likely because the AWG (arbitrary waveform generator using an 8-bit digital to analogue converter) drive signal frequency was  6 GHz offering 12 GSa/s whereas the SOA used had a -3dB  frequency response of 0.6 GHz, therefore we over-sampled the drive signal by approximately ~10x. In a real DCN scenario, to implement our algorithms' driving signals in practice, we would likely use an FPGA or ASIC with an embedded on-chip DAC  for multilevel signal generation, and there are already existing FPGAs (a.k.a. RF System on Chip (RFSoC)) that  support multiple DACs at 6 GSa/s. Therefore in practice the search space would be lower (fewer dimensions/number of points to optimise) than assumed in this paper, and we would expect this to improve the AI convergence characteristics. Further experiments using fewer points in the drive signal/a slower AWG are necessary to see what the true effects are on the AI algorithms. This is beyond the scope of this paper, and we intend to further investigate it in our future work.

Within the context of a DCN implementation of the presented methods, some considerations were made with respect to the effect that the algorithms have on the signal to noise ratio (SNR). Namely, it should be considered if the oscillations caused by the algorithms (all of which are of the order of 5\%) have a negative effect on the SNR of the ON period of the output, particularly in comparison to the output of a step driving signal, where the ON period considered is defined as starting when the signal enters the $\pm 5\%$ (with respect to the steady state) region for a 20 ns pulse length. Following from the model of amplifier noise given in \cite{agrawal} and accounting for Shot noise, intrinsic amplifier noise (the noise figure of the SOA) and the additional noise due to the fluctuations in the output, we consider the penalty on the noise figure (as defined in \cite{agrawal}) due to the deviations of the output from its steady state value throughout the duration of its ON period. Assuming (based on intrinsic and Shot noise contributions) a base noise figure (i.e. if the driving method caused no deviations at all) of 7.1dB, the measured noise figure penalties for ACO, PSO, GA and step were 1.05 dB, 0.65 dB, 1.12 dB and 0.53 dB with SNR values of 28.52 dB, 28.90 dB, 28.54 dB and 29.06 dB respectively, showing that the additional noise figure penalty due to the AI methods ranges between  0.08 dB (PSO) and 0.59 dB (GA) compared to a step in the case of the best performing algorithm (PSO).

\section{Conclusion} 

%
%
%
%





Simulation and experimental results of SOA off-on switching were presented for various driving signal formats. The paper outlined a novel approach to SOA driving signal generation with AI algorithms which made no assumptions about the SOA and therefore were general, required no historic data collection and could be scaled to any SOA-based switch, opening up the possibility of rapid all-optical switching in real data centres. Experimental settling times (and therefore effective off-on times) of 547 ps were achieved using PSO, offering an order of magnitude performance improvement with respect to settling time over our implementation of the PISIC and MISIC techniques from the literature. Additionally, the standard PID control and raised cosine techniques from control theory were shown to be inadequate for the problem of ultra-fast SOA switching. Although ACO and GA demonstrated slightly faster rise times than PSO, PSO had a faster settling time and also a significantly lower 1.8\% cost spread, giving greater reliability that any given PSO run had found the optimum solution. Furthermore, due to the fewer restrictions placed on the search space and the lower number of fine-tuned hyperparameters compared to ACO and GA, PSO was found to be more easy to generalise to unseen SOAs. Future work expanding on the presented methods could examine the robustness of the method with respect to hardware limitations/irregularities (e.g. temperature/bias current variations, bit resolution of driving voltage values or the sampling frequency of the AWG). Additionally, future work could extend the method to a scenario consisting of multiple, possibly cascaded, SOAs.

\bibliographystyle{ieeetr}
\bibliography{bibliography}








\end{document}